\documentclass[twocolumn,aps,prx,superscriptaddress]{revtex4-2}
\usepackage[T1]{fontenc}
\usepackage{babel}
\usepackage{amsmath}
\usepackage{amsfonts}
\usepackage{amssymb}
\usepackage{amsthm}
\usepackage{bbm}
\usepackage{dsfont}
\usepackage{graphicx}
\usepackage{braket}
\usepackage{multirow}
\usepackage[unicode=true]{hyperref}
\usepackage{xcolor}
\usepackage{yfonts}
\usepackage{bm}
\usepackage{mathrsfs}
\usepackage{graphicx}
\usepackage{verbatim}
\usepackage{hyperref}
\usepackage{wasysym}
\usepackage[normalem]{ulem}

\begin{document}

\preprint{APS/123-QED}

\title{Thermal entanglement transitions from strong $\textrm{SU}(2)$ symmetry}

\author{Siqi Mo}
\affiliation{Department of Physics, University of California, Berkeley, CA 94720, USA}
\author{Ehud Altman}
\affiliation{Department of Physics, University of California, Berkeley, CA 94720, USA}
\affiliation{Materials Sciences Division, Lawrence Berkeley National Laboratory, Berkeley, CA 94720, USA}
\author{Samuel J. Garratt}
\affiliation{Department of Physics, Princeton University, Princeton, NJ 08544}
\affiliation{Department of Physics, University of California, Berkeley, CA 94720, USA}


\begin{abstract}

We study mixed-state entanglement in Gibbs states with strong $\mathrm{SU}(2)$ symmetry, focusing on locally interacting spin systems with ferromagnetic interactions. Our main result is to show that finite-temperature ordering transitions are associated with entanglement transitions, and therefore the steady state of strongly symmetric Lindbladians can exhibit entanglement transitions. While in the paramagnetic phase it is known that the distillable entanglement and logarithmic entanglement negativity between two halves of a large system grow logarithmically with the square root of the number of spins, we show that in ferromagnetic phases these quantities grow parametrically faster with system size. To arrive at this result we first establish relations between these mixed-state entanglement measures and spin correlations in states that are singlets under global $\rm{SU}(2)$ symmetry transformations. We then introduce a semiclassical theory for $\rm{SU}(2)$ singlet thermal states. While the global singlet constraint generally enters this theory as a complicated function of the full semiclassical spin configuration, we show that in large $S$ limit it simplifies drastically to a Gaussian suppression of total magnetization in disordered phases as well as in ordered phases in the vicinity of continuous thermal phase transitions. This leads us to a field theory describing spin correlations in the singlet sector. Using this we determine the behavior of various probes of mixed-state entanglement at low and at high temperatures, supporting our analytical results using numerical Monte Carlo simulations of a three-dimensional lattice realization of our semiclassical theory. We also use exact numerics in one-dimensional spin-$1/2$ systems to confirm our predictions for the scaling of mixed-state entanglement with correlation length and system size in the paramagnetic phase. 
\end{abstract}

\maketitle

\section{Introduction}

Repeated interactions between quantum systems and their environments typically degrade entanglement within the system. Such interactions drive systems toward states having the maximum possible entropy subject to any physical constraints imposed by the environment. For example, if the environment exchanges both energy and particles with the system, the late-time state is a grand canonical thermal ensemble, whereas if the number of particles in the system is conserved then one finds a canonical ensemble, having a strong $\rm{U}(1)$ symmetry. Both kinds of ensemble appear to be only weakly entangled at finite temperature~\cite{Hasting2011FiniteTO,Sherman2016SpinNegativity,Hart2018TCNegativity,Bakshi2025SuddenDeath,Garratt2025Private,Negari2025AbelianHigh,kuwahara2025clustering,scalet2026spatial}. 

In stark contrast to this behavior, Refs.~\cite{Li2025HighlyEntangled, Moharramipour2024MMIS} recently showed that infinite-temperature states with strong non-Abelian symmetries have bipartite distillable entanglement that diverges with system size. Such states arise when system-environment interactions are invariant under non-Abelian symmetry transformations of the system Hamiltonian. However, it is unclear whether and how this behavior is modified at low temperatures. 

In this work we show that entanglement transitions can occur as the temperature is decreased. Our focus is on spin systems with $\rm{SU}(2)$ symmetry, and on the structure of thermal states in the total singlet sector, which are strongly symmetric under all global $\rm{SU}(2)$ transformations. Throughout this work we refer to these as singlet Gibbs states. We derive general relations between measures of mixed-state entanglement and spin correlations in the singlet sector, and show how these correlations can be understood from a semiclassical theory that is appropriate for large microscopic spin $S \gg 1$ and in large systems. This semiclassical theory takes a simple form at high temperatures and in the vicinity of continuous thermal phase transitions. Using this theory we show that finite-temperature ferromagnetic ordering transitions are associated with transitions in the scaling of various mixed-state entanglement measures with system size. 

\begin{figure}
\includegraphics[width=\linewidth]{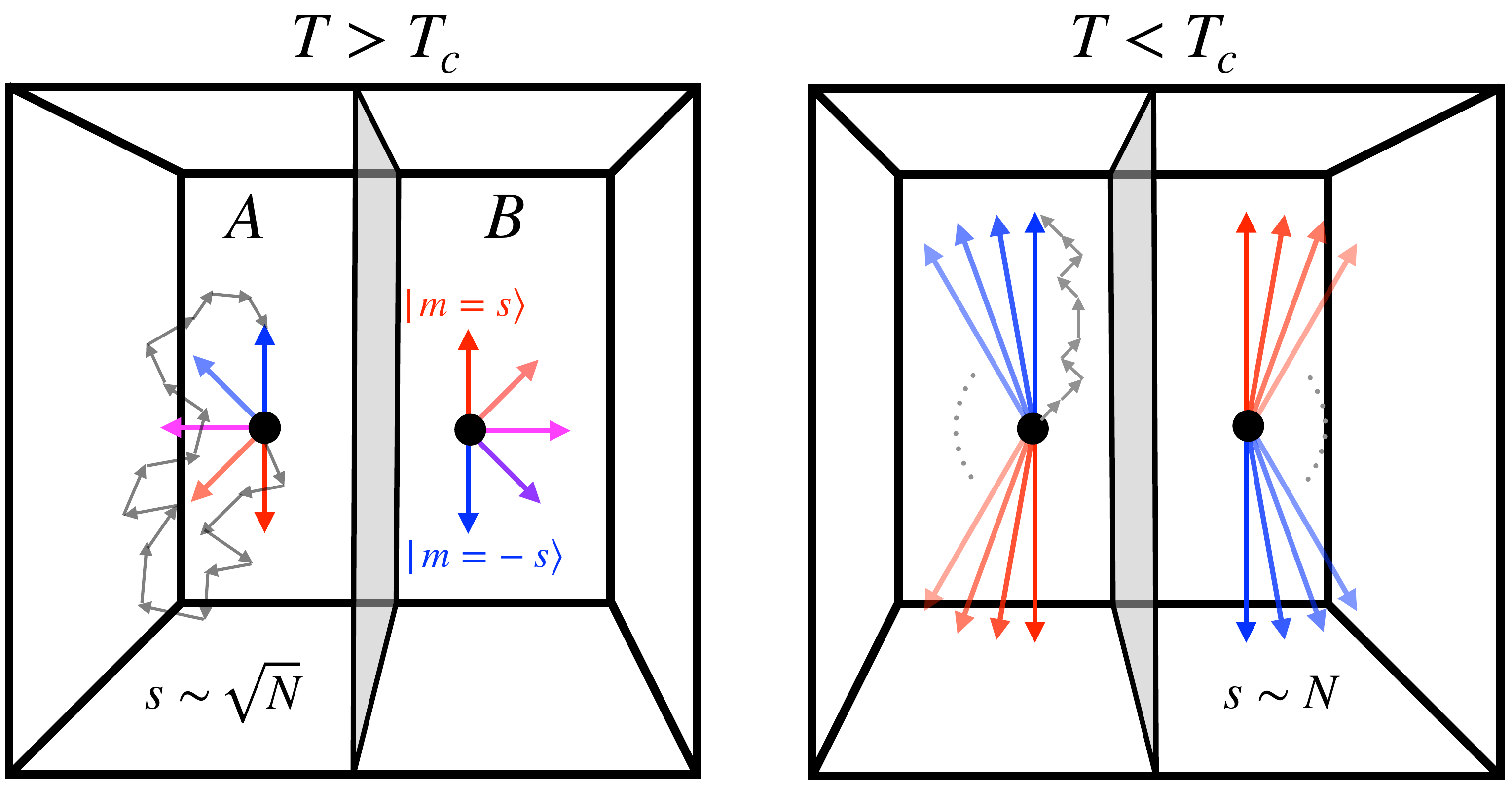}
\caption{Coincidence of thermal and entanglement transitions in the singlet sector of locally interacting ferromagnets. Due to the singlet constraint, pure states with half-system spin $s$ are entangled superpositions with Schmidt rank $2s+1$. Each of the two subsystem $A$ and $B$ contains approximately $N/2$ spins and these must be entangled into a total singlet. In the paramagnetic phase the half-system spin (colored arrows) scales as $s \sim \sqrt{N}$, which semiclassically can be understood as arising from a random walk of $N/2$ steps in spin space (grey arrows). In the ferromagnetic phase, and with anisotropy that favors anti-aligned $A$ and $B$, we instead have $s \sim N$, although this scaling depends sensitively on system geometry as we discuss. The transition in the scaling of $s$ with $N$ that occurs at the critical temperature $T_c$ is therefore associated with a transition in mixed-state entanglement between $A$ and $B$.}
\label{fig:randomwalkdiagram}
\end{figure}

This behavior should be contrasted with that of nonlocal models featuring all-to-all interactions between spins. While at infinite temperature and across an equal bipartition, the distillable entanglement scales as $\log \sqrt{N}$ up to subleading corrections \cite{Moharramipour2024MMIS,Li2025HighlyEntangled}, at finite-temperature and in a nonlocal model Ref.~\cite{Moharramipour2024MMIS} showed that this quantity is only of order unity. Our semiclassical theory, as well as exact numerics in one spatial dimension, reveal a very different although rather intuitive picture of finite-temperature entanglement in the singlet sectors of local models. For example, for ferromagnetic Hamiltonians, low-energy states can be constructed by separately aligning the spins in two halves of the system, and then pairing the two resulting macroscopic spins into a singlet. This singlet, being formed from macroscopic spins whose magnitudes are proportional to system volume, has distillable entanglement $\log N$, up to corrections that are subleading at large $N$. This situation is illustrated in Fig.~\ref{fig:randomwalkdiagram}. Although this particular state is only a caricature of one of the contributions to a low-temperature singlet Gibbs state, a related picture will emerge from our theory. Interestingly we will see that in various geometries and in the ordered phase, Goldstone fluctuations modify the scaling of distillable entanglement. Throughout the finite-temperature paramagnetic phase of local models, we will find distillable entanglement $\log \sqrt{N}$ (up to subleading additive corrections) as at infinite temperature. This is consistent with very recent work demonstrating $\log\sqrt{N}$ scaling in one spatial dimension \cite{Zeng_2026_Thermal}.

Decreasing the temperature in systems with strong Abelian symmetries has milder consequences for mixed-state entanglement than the non-Abelian case studied here. First note that, while conventional thermal states of locally interacting qudit systems are separable above a finite threshold temperature \cite{Bakshi2025SuddenDeath}, this is no longer the case when there is a strong symmetry, in which case there is generically entanglement at any finite temperature \cite{Garratt2025Private,Negari2025AbelianHigh, Kim2025TopoNegativity}. Moreover, when the temperature is decreased below a symmetry-breaking phase transition, strongly symmetric thermal states become long-range entangled, in the sense that any representation as a statistical mixture of pure states must involve pure states with divergent circuit complexity \cite{Lu2023LongRange,Garratt2025Private,McGinley2025Lower}. That said, in ordered phases with spontaneously broken Abelian global symmetries, the distillable entanglement and logarithmic negativity across a bipartition are independent of system size. As discussed above, for $\rm{SU}(2)$ symmetric systems these measures of mixed-state entanglement are instead parametrically large in system size. 

Part of the motivation for this work comes from recent breakthroughs in the design~\cite{Chen2025GibbsPrep, Ding2025GibbsPrep, Ding2025GibbsPrep2, Lloyd2025ThermalPrep, Song2025VariCool,Chen2025efficient,Hahn2026efficient}  and implementation~\cite{Mi2024Google} of protocols for thermal state preparation. Quantum Gibbs samplers, which are quantum channels or Lindbladians having thermal steady states, can be implemented by simulating the interaction between a system and the environment. Restricting the form of the interaction, as well as the initial state of the system, is expected to prepare a symmetrized thermal ensemble as discussed above. Specifically, by starting from a $\mathrm{SU}(2)$ singlet initial state (for example, a nearest-neighbor dimer covering of the lattice), and enforcing the Lindbladian jump operators to be $\mathrm{SU}(2)$ symmetric, the resulting steady state is the singlet Gibbs state. Therefore, our result implies that there are fairly simple open-system evolution operators whose steady states undergo sharp entanglement transitions when operator parameters (e.g. the temperature) are tuned. 

In practice, generating long-range mixed-state entanglement using a quantum Gibbs sampler involves local measurements and local feedback. This route should be contrasted with strategies where feedback operations are chosen based on nonlocal functions of measurement outcomes \cite{Hastings2021dynamically,Tantivasadakarn2023hierarchy,Zhu2023nishimori,Lu2023LongRange}. Moreover, because thermal state preparation is based simply on the maximization of entropy (subject to global constraints) we anticipate that our route does not depend strongly on details of the interactions within the system or with the environment. The basic requirement is simply that the system is able to thermalize.

Our approach for identifying the entanglement transition has two key steps. The first, which extends an argument of Ref.~\cite{Moharramipour2024MMIS,Li2025HighlyEntangled}, is to relate measures of mixed-state entanglement to correlations between spins in the singlet Gibbs state. The second, more technical step, is to derive a semiclassical theory for spin correlations in the singlet sector. In this theory spin correlations are expressed as averages over contributions from different spin coherent states; the central problem is to capture the effect of the singlet constraint on such averages. The singlet constraint has an extremely complicated form in general, being a nonlocal function of the configuration of all of the spins. However, we show that this constraint drastically simplifies in disordered phases as well at high temperatures in ordered phases close to the phase transition, provided the thermal phase transition separating the ordered and disordered phases is continuous. 

In these regimes we show that the effect of the singlet constraint on spin correlations can be captured simply by a Gaussian suppression of the total semiclassical spin. We perform numerical calculations to show that, for typical spin coherent states, this Gaussian approximation is accurate for large numbers $N$ of spins and for arbitrary local spin $S$. While focusing on large $S$ will simplify our analysis of spin correlations, we do not believe that this choice significantly modifies the behavior. Using a Monte Carlo calculation of semiclassical spin correlations in a three-dimensional Heisenberg ferromagnet, and the Gaussian approximation to the singlet constraint, we then verify that there is a finite-temperature transition in the properties of the spin configuration that provide lower bound on mixed-state entanglement.  


This work is organized as follows. First, in Sec.~\ref{sec:structureofsinglet} we define the singlet Gibbs state and discuss the structure of the singlet sector of many-body Hilbert space for spin systems. In Sec.~\ref{sec:entanglement} we relate the lower bounds on distillable entanglement and the logarithmic negativity to correlations between the spins for generic $\mathrm{SU}(2)$ singlet states. In Sec.~\ref{sec:decoupled} we study a simple model for an entanglement transition where the Hamiltonian is decoupled across the entanglement cut. In Sec.~\ref{sec:SCT} we develop a semiclassical theory that describes spin correlations in the presence of the singlet constraint, and show that the finite-temperature ferromagnetic phase transition coincides with a mixed-state entanglement transition. In Sec.~\ref{sec:numerics} we numerically investigate one-dimensional systems, showing that the divergence of the correlation length is associated with an increase in mixed-state entanglement.  

\section{Structure of the singlet sector}\label{sec:structureofsinglet}

In this work we study mixed-state entanglement in $\rm{SU}(2)$ symmetric spin systems, focusing on the case where the interactions are local and ferromagnetic. Our focus is on entanglement in the Gibbs state restricted to the total singlet sector, and we describe this object below. 

We consider a lattice with a spin $S$ on every site and denote by ${\vec{S}}_j=({S}^x_j,{S}^y_j,{S}^z_j)$ the vector of spin operators at site $j=1,\ldots,N$, where $N$ is the total number of spins. We denote the total spin as ${\vec{S}}_{\mathrm{tot}}\equiv \sum_{j=1}^N {\vec{S}}_j$. Global $\rm{SU}(2)$ rotations are represented by unitary operators $U^{\otimes N}$,
\begin{align}
    U^{\otimes N} \equiv \bigotimes_{j=1}^N e^{i \gamma \hat n\cdot {\vec{S}}_j} = e^{i \gamma \hat n\cdot {\vec{S}}_{\mathrm{tot}}}.
\end{align}
The operator $U^{\otimes N}$ is an implicit function of the rotation angle $\gamma$ and axis $\hat n$ (a three-component unit vector). 

We consider a Hamiltonian $H$ which describes geometrically local interactions between spins. $H$ has global $\rm{SU}(2)$ symmetry $[U^{\otimes N}]H [U^{\otimes N}]^{-1}=H$ and, due to this symmetry, it can be block diagonalized into blocks labeled by eigenvalues of the squared total spin ${\vec{S}}^2_{\mathrm{tot}}$ and of ${S}^z_{\mathrm{tot}}$. We denote the projector onto the total singlet sector, where ${\vec{S}}^2_{\mathrm{tot}} = {S}^z_{\mathrm{tot}} = 0$, by $\Pi_0$. The singlet Gibbs state at inverse temperature $\beta$ is defined as
\begin{equation}
	\rho_0 = \frac{1}{\mathcal{Z}} \Pi_0 e^{-\beta H},
\end{equation}	 
where $\mathcal{Z}=\mathrm{Tr}\left[\Pi_0 e^{-\beta H}\right]$. By construction this density matrix is strongly symmetric with respect to all global $\rm{SU}(2)$ rotations $U^{\otimes N}\rho_0=\rho_0 U^{\otimes N}=\rho_0$.

We study the entanglement between a subsystem $A$ and its complement $B$, each making up approximately half of the system. Since the interactions are geometrically local we can write
\begin{align}
    H = H_A + H_B + H_{AB},
\end{align}
where $H_A$ ($H_B$) acts only on degrees of freedom in $A$ ($B$) and $H_{AB}$ describes the interactions between $A$ and $B$. For a $d$-dimensional geometrically local system with linear extent $L$, $H_{AB}$ consists of $\mathcal{O}(L^{d-1})$ terms along the boundary of $A$ and $B$. Importantly, the three sets of terms $H_A$, $H_B$ and $H_{AB}$ are individually symmetric under $\rm{SU}(2)$. 

A convenient basis for the singlet sector comes from entangling sets of degenerate eigenstates of $H_A$ and $H_B$. Because $H_A$ is $\mathrm{SU(2)}$ symmetric, its eigenstates can be labeled by $\ket{s_A, m_A, a}_A$, where $s_A(s_A+1)$ is the eigenvalue of the squared spin operator ${\vec{S}}_A^2$ for subsystem $A$, explicitly ${\vec{S}}_A^2 = (\sum_{j \in A} {\vec{S}}_j)^2$, and where $m_A$ is the eigenvalue of ${S}^z_A \equiv \sum_{j \in A} {S}^z_j$. The index $a$ labels distinct $(2s_A+1)$-fold degenerate multiplets in the total spin-$s_A$ sector of subsystem $A$, and $m_A=-s_A,-s_A+1,\ldots,s_A$ labels states within a multiplet, e.g. $\ket{s_A,m_A-1, a}_A \propto {S}^-_A\ket{s_A,m_A,a}_A$ with ${S}^-_A$ the spin lowering operator for subsystem $A$. Eigenvalues of $H_A$ are denoted $E_A(s_A,a)$ with $H_A\ket{s_A,m_A,a}_A=E_A(s_A,a)\ket{s_A,m_A,a}_A$. The basis $\ket{s_B,m_B,b}_B$ of subsystem $B$ is constructed similarly. 

An orthonormal and complete basis for the entire system $AB$ in the total singlet sector with ${\vec{S}}_{\text{tot}}^{2}=0$ can be constructed by superposing $\ket{s_A,m_A,a}_A \otimes \ket{s_B,m_B,b}_B$ over $m_A, m_B$ in a way that obeys the singlet coupling rule, which requires $s_A = s_B \equiv s$ and $m_A = -m_B \equiv m$, symmetrized properly over $m$. This gives 
\begin{equation}\label{singbasis}
    \ket{s,a,b} = \frac{1}{\sqrt{2s+1}}\sum_{m=-s}^{+s}(-1)^{m} \ket{s,m,a}_A \otimes \ket{s,-m,b}_B.
\end{equation}
Such states satisfy ${\vec{S}}_{\text{tot}}^2\ket{s,a,b}=0$ as well as
\begin{equation}\label{eq:basis_properties}
\begin{split}
        {\vec{S}}_A^2\ket{s,a,b}&={\vec{S}}_B^2\ket{s,a,b}=s(s+1)\ket{s,a,b},\\
    	H_A\ket{s,a,b}&=E_A(s,a)\ket{s,a,b}, \\ H_B\ket{s,a,b}&=E_B(s,b)\ket{s,a,b}.
\end{split}
\end{equation}
The above basis states $\ket{s,a,b}$ can be intuitively understood as spin-$s$ analogs of EPR states, and will play an important role in the next section, where we connect mixed-state entanglement in $\rho_0$ to spin correlations.

\section{Entanglement of Gibbs state in the singlet sector}\label{sec:entanglement}

The aim of this section is to derive operationally meaningful expressions for mixed-state entanglement measures in the singlet Gibbs state and relate them to physical observables, specifically subsystem total spin in this state. The relations that we derive here generalize immediately to any singlet state.

\subsection{Lower Bound on Distillable Entanglement}\label{subsec:Distillable_lowerbound}

The distillable entanglement between subsystems $A$ and $B$ of a state $\rho$, denoted by $E^D_{A:B}(\rho)$, is defined as the maximum asymptotic rate at which standard Einstein-Podolsky-Rosen (EPR) pairs can be distilled from infinitely many copies of $\rho$, through local operations and classical communication (LOCC) performed by the two parties $A$ and $B$ \cite{Bennett1996Distillabel}. Here ``maximum'' refers to the optimal LOCC protocol. 
Besides having this operational meaning, the distillable entanglement provides lower bounds on a wide variety of other mixed-state entanglement measures satisfying standard axioms (non-negativity, vanishing for separable states, normalization, monotonicity under LOCC, convexity, continuity, and partial additivity) \cite{Horodecki2000Limits}. Moreover, \emph{any} LOCC protocol performed by parties $A$ and $B$ on state $\rho$ can provide us with a lower bound on $E^D_{A:B}(\rho)$ --- below we will construct such a lower bound by considering a specific entanglement distillation protocol. The state $\rho$ of interest will almost always be the singlet Gibbs state $\rho_0$, so we write $E^D_{A:B} \equiv E^D_{A:B}(\rho_0)$ for brevity.

The idea is that the two parties $A, \, B$ perform operations to distill one of the pure basis states $\ket{s,a,b}$ [Eq.~\eqref{singbasis}] from the singlet Gibbs state $\rho_0$. Recall that the individual $\rm{SU}(2)$ symmetries of $H_A$ and $H_B$ enforce their eigenstates to come in degenerate multiplets. So we first let party $A$ measure $\vec{S}_A^2$ and $H_A$: by finding their eigenvalues $s(s+1)$ and $E_A(s,a)$, party $A$ has applied a projection operator
\begin{align}
	\Pi_{s,a}^A = \sum_{m=-s}^s \ket{s,m,a}_A \bra{s,m,a}
\end{align}
to subsystem $A$. In this measurement the observed outcomes, which we can label with $s$ and $a$, have probabilities $\text{Tr}\left[\rho_0 \Pi_{s,a}^A\right]$. We then let party $B$ measure $H_B$, so they apply to subsystem $B$ the projector
\begin{align}
	\Pi_{s,b}^B = \sum_{m=-s}^s \ket{s,m,b}_B\bra{s,m,b}.
\end{align}
We note that because $\rho_0$ is a total singlet, the eigenvalue of ${\vec{S}}_B^2$ is \emph{fixed} to match the observed eigenvalue of ${\vec{S}}_A^2$. That is, parties $A$ and $B$ must both observe the same `$s$'. Here we have assumed that $E_A(s,a)$ and $E_B(s,b)$ for different $a$ and $b$, respectively, are non-degenerate, so the post-measurement state is a pure state $\ket{s,a,b}$ --- this assumption is valid in thermodynamically large system without any special structure in $H_A, H_B$ where accidental degeneracies are non-generic. But more generally we can simply imagine that parties $A$ and $B$ perform measurements whose outcomes correspond to the projectors above. The result of the two measurements is that the entangled state $\ket{s,a,b}$ is created with probability 
\begin{align}
	p(s,a,b) = \text{Tr}\left[\Pi_{s,a}^A\Pi_{s,b}^B \rho_0 \right].\label{eq:p_sab}
\end{align}

Once parties $A$ and $B$ communicate with one another, they know exactly which pure state they have created. Because $\ket{s,a,b}$ is maximally entangled, they distill $\log_2(2s+1)$ bits of entanglement \cite{Bennett1996Distillable} with probability $p(s,a,b)$. The distillable entanglement is therefore lower bounded by the distillation rate of this protocol:
\begin{align}
	E^D_{A:B} \geq \sum_{s}p(s) \log_2(2s+1),\label{eq:distillablelowerbound0}
\end{align}
where $p(s)=\sum_{ab}p(s,a,b)$. This lower bound, which we denote by $E^{D\;<}_{A:B}$, can equivalently be written as
\begin{align}
	E^{D\;<}_{A:B} = \mathrm{Tr}\left[ \rho_0 \log_2\sqrt{1 + 4 \vec{S}_A^2 }\right],
    \label{eq:distillablelowerbound}
\end{align}
and is a direct relation between half-system total spin and distillable entanglement.

Note that if $\rho_0$ is diagonal in the singlet basis $\ket{s,a,b}$ the above protocol is optimal \cite{Sahu2025EntanglementHierarchies}. Additionally, in that case, all mixed-state entanglement measures satisfying the aforementioned axioms match Eq.~\eqref{eq:distillablelowerbound}. This is the case at infinite temperature \cite{Li2025HighlyEntangled,Moharramipour2024MMIS} and, as we discuss in Sec.~\ref{sec:decoupled}, at all temperatures for Hamiltonians with no interactions across the entanglement cut $H_{AB}=0$. 

\subsection{Lower Bound on Logarithmic Negativity} \label{subsec:negativity_lowerbound}

Here we derive a related lower bound on the logarithmic entanglement negativity, defined as \cite{Vidal2001Negativity}
\begin{align}
	E^N_{A:B}(\rho) \equiv \log_2 ||\rho^{T_A}||_1,
    \label{eq:logneg}
\end{align}
where $\rho^{T_A}$ is the partial transpose (in subsystem $A$) of the density matrix $\rho$ , and $||\cdots||_1$ is the $1$-norm (the sum of singular values). 

While the logarithmic entanglement negativity is not strictly an axiomatic mixed-state entanglement measure, due to non-convexity \cite{Plenio2005Negativity}, it is related to entanglement cost under quantum operations preserving the positivity of partial transpose \cite{Audenaert2003PPTEntanglement}. It also satisfies $E^{D}_{A:B}(\rho) \leq E^N_{A:B}(\rho)$ \cite{Vidal2001Negativity}. Important for our purposes is the fact that logarithmic entanglement negativity is an entanglement monotone, meaning that it is non-increasing under LOCC \cite{Plenio2005Negativity}. Considering the POVM 
\begin{align}
	\mathcal{E}(\rho) = \sum_{s_A s_B ab} \Pi^A_{s_A,a} \Pi^B_{s_B,b} \rho \Pi^A_{s_A,a} \Pi^B_{s_B,b}, \label{eq:POVM}
\end{align}
where $\rho$ is a general density matrix of $AB$, we therefore have $E^N_{A:B}(\mathcal{E}(\rho)) \leq E^N_{A:B}(\rho)$. Physically this POVM corresponds to the measurement protocol described above, followed by discarding the observed outcomes $s_A, s_B,a,b$. 

Applying the channel $\mathcal{E}$ to the singlet Gibbs state $\rho_0$ we find
\begin{align}
	\mathcal{E}(\rho_0) = \sum_{sab} p(s,a,b) \ket{s,a,b}\bra{s,a,b},
\end{align}
where $p(s,a,b)=\mathrm{Tr}\left[\Pi^A_{s,a}\Pi^B_{s,b}\rho_0\right]$ as above, and we note again that here $s_A=s_B$ since $\rho_0$ is a singlet. For this state it is straightforward to evaluate the logarithmic negativity,
\begin{align}
	E^N_{A:B}(\rho_0) \geq E^N_{A:B}(\mathcal{E}(\rho_0)) &= \log_2\left[\sum_{s} p(s)(2s+1)\right].\label{eq:negativitylowerbound0}
\end{align}
Rewriting the right-hand side as the expectation of an operator, we arrive at our lower bound
\begin{align}
	E^{N\;<}_{A:B} = \log_2\mathrm{Tr}\left[ \rho_0 \sqrt{1+4\vec{S}_A^2}\right], \label{eq:negativitylowerbound}
\end{align}
which also relates mixed-state entanglement to half system total spin. 

Although the entanglement negativity $E^N_{A:B}$ is itself an upper bound on the distillable entanglement $E^D_{A:B}$, there is no definite general relation between our lower bound in Eq.~\eqref{eq:negativitylowerbound} and $E^D_{A:B}$. We will nevertheless identify transitions in the lower bound as the temperature is varied. 

\section{Decoupled Hamiltonian}\label{sec:decoupled}
As a first step, we show that the entanglement transition coincides with a finite-temperature ordering transition in a decoupled model with $H = H_A+H_B$ and $H_{AB}=0$. In particular we consider a setting where $A$ and $B$ are identical decoupled subsystems such that $H_A$ and $H_B$ share the same $\mathrm{SU}(2)$ symmetric form.
 
As noted previously in Eq.~\eqref{eq:basis_properties}, the singlet energy eigenstates of this decoupled Hamiltonian take the form of $\ket{s,a,b}$ in Eq. \eqref{singbasis}, and their eigenvalues with respect to $H$ are simply $E(s,a,b) = E_A(s,a) + E_B(s,b)$. The singlet Gibbs state takes the form
\begin{equation} \label{eq:decoupledrho}
\begin{split}
    \rho_0 =  \frac{1}{\mathcal{Z}}\sum_{s,a,b} e^{-\beta [ E_A(s,a) + E_B(s,b)]} \ket{s,a,b} \bra{s,a,b}.
\end{split}
\end{equation}

An important feature of this $\rho_0$ is that it is diagonal in the singlet basis. As noted in the previous section, $E^{D}_{A:B}$ and all other mixed state entanglement measures satisfying the aforementioned axioms saturate to the lower bound given in Eq.~\eqref{eq:distillablelowerbound}, and in this case they identically take the form of 
\begin{equation}
    E^{D}_{A:B} = \sum_{s} p(s)\log_2(2s+1),
\label{eq:EDdecoupled}
\end{equation}
where for the decoupled model the probability distribution $p(s) \propto \mathcal{Z}_A^2(s)$, with $\mathcal{Z}_A(s)=\sum_a e^{-\beta E_A(s,a)}$ the partition function for subsystem $A$ restricted to have total spin $s$. We have used the fact that the two subsystems and Hamiltonians are identical. 

Another important feature of the decoupled model is that its singlet Gibbs state is invariant under the POVM in Eq.~\eqref{eq:POVM}. This means that the logarithmic entanglement negativity is equal to the expression in Eq.~\eqref{eq:negativitylowerbound},
\begin{equation}
   	E^N_{A:B}(\rho_0) = \log_2\left[\sum_s p(s)  (2s+1) \right],\label{eq:negativitydecoupled}
\end{equation}
i.e. in this case $E^N_{A:B}(\mathcal{E}(\rho_0))=E^N_{A:B}(\rho_0)$. This expression is similar to $E^{D}_{A:B}(\rho)$ except the average is inside the logarithm, consistent with the relation $E^N_{A:B}(\rho_0) \ge E^{D}_{A:B}(\rho_0)$. 

To understand the behavior of $E^{D}_{A:B}$ and $E^N_{A:B}$ for this model we consider the moments of $\vec{S}_A^2$ in the standard (weakly symmetric) Gibbs state of subsystem $A$, which is proportional to $e^{-\beta H_A}$. In this state we can express this expectation value as
\begin{align}
	\frac{\text{Tr}[e^{-\beta H_A} \vec{S}_A^2]}{\text{Tr}[e^{-\beta H_A}]} = \sum_s p_A(s) s(s+1),
    \label{eq:SA2}
\end{align}
where ${p_A(s) \propto \mathcal{Z}_A(s)(2s+1)}$ with $\sum_s p_A(s)=1$.  We note also that since $\vec{S}_A^2 = \sum_{jk \in A} \vec{S}_j \cdot \vec{S}_k$ is a sum of bounded operators, the central limit theorem applies in both of the phases (although not necessarily at the critical point). The implication is that $p_A(s)$ has a distinct maximum rather than, for example, a heavy tail at large $s$. 

We focus on the case where $H_A$ and $H_B$ are nearest-neighbor ferromagnetic Heisenberg model in spatial dimension $d \geq 3$. The mean of $p_A(s)$ in the standard Gibbs state of $H_A$ undergoes a transition from $s \sim N^{1/2}$ to $s \sim N$ as the temperature is lowered through the paramagnet-to-ferromagnet transition. To see that the transition in the singlet Gibbs state is unchanged note
that the relation
$p(s)\propto (2s+1)^{-2}p^2_A(s)$ implies that the free energy $F(s) \equiv -\ln p(s)$ of sector $s$ in the singlet Gibbs state can be expressed in terms of the free energy of an unrestricted subsystem $F_A(s) = -\ln p_A(s)$ as
$F(s) = 2F_{H_A}(s) + 2\ln(2s+1) + \ldots$, where the ellipsis denotes $s$-independent contributions. 
The addition of $\ln(2s+1)$ is subextensive and therefore cannot modify a phase transition driven by a competition between energy and entropy, each of which are extensive. The logarithmic entanglement negativity, here given by Eq.~\eqref{eq:negativitydecoupled} will therefore transition from $\frac{1}{2}\log N$ to $\log N$ on passing through the phase transition, where we have neglected corrections that are subleading at large $N$.


The usual expectation is that re-introducing the coupling $H_{AB}$ across the $(d-1)$-dimensional $AB$ boundary would not alter thermodynamic phases and phase transitions. However, mixed-state entanglement measures are intrinsically non-local quantities, and are known to be sensitive to other effects that do not alter thermodynamics, such as global symmetry constraints \cite{Garratt2025Private,Negari2025AbelianHigh}. In the next section we will indeed find that, for $H_{AB} \neq 0$, probes of mixed-state entanglement in the ordered phase can differ from the simple $\log N$ scaling identified above. 

\section{Semiclassical theory of singlets}\label{sec:SCT}

In this section we develop a semiclassical description of the singlet Gibbs state and then apply it to bound the behavior of the distillable entanglement and the logarithmic entanglement negativity. For concreteness we focus on Heisenberg ferromagnets, with Hamiltonian $H=-\frac{J}{S^2}\sum_{\langle jk\rangle}{\vec{S}}_j \cdot {\vec{S}}_k$, but our approach is more general.

\subsection{Singlet constraint}
\label{subsec:smiclass}
The singlet Gibbs state density matrix can be represented as
\begin{equation}\label{eqn:rhosym}
\begin{split}
    \rho_{0} & \propto \Pi_0e^{-\beta H} \propto \int dU U^{\otimes N}e^{-\beta H},
\end{split}
\end{equation}
where $U^{\otimes N}$ is a tensor product of identical $\mathrm{SU}(2)$ rotations $U$ acting on each of the $N$ individual spins, and $dU$ is the Haar measure for $\mathrm{SU}(2)$. The partition function $\mathcal{Z}_0 = \text{Tr}[\Pi_0 e^{-\beta H}]$ is then expressed as a spin coherent state path integral
\begin{equation}\label{eqn:Zsym}
\begin{split}
   \mathcal{Z}_0 \propto \int dU \int D \bm{\hat{\Omega}}(\tau) e^{-\mathcal{A}[\bm{\hat{\Omega}}(\tau)]} \bra{\bm{\hat{\Omega}}(0)} U^{\otimes N}\ket{\bm{\hat{\Omega}}(\beta)},
\end{split}
\end{equation}
where $\hat{\Omega}_j(\tau)$ is a three-component unit vector associated with site $j$ and imaginary time $\tau \in [0,\beta]$, $\bm{\hat{\Omega}}(\tau)$ is the full spin configuration at $\tau$, and the integration measure for the spins $D \bm{\hat{\Omega}}(\tau)=\prod_j D\hat{\Omega}_j(\tau)$. The spacetime action, at generic $S$ and without the singlet constraint, is
\begin{equation}\label{eqn:SHeis}
    \mathcal{A}[\hat{\bm{\Omega}}]=-\int_0^{\beta} d\tau\; J\sum_{\langle jk\rangle} \hat{\Omega}_j(\tau) \cdot \hat{\Omega}_k(\tau) - iS \sum_j \omega(\hat{\Omega}_j),
\end{equation}
where $iS \omega(\hat{\Omega}_j)$ is the geometric phase term which depends on the full history $\bm{\hat{\Omega}}(\tau)$. Note that the prefactor of the energetic term is independent of $S$ because we have chosen the quantum Hamiltonian to take the form $H = -(J/S^2)\sum_{\langle j k \rangle}\hat{S}_j \cdot \hat{S}_k$. 

At large $S$, $\tau$-dependent histories $\bm{\hat{\Omega}}(\tau)$ are suppressed by the rapid oscillations from the $iS \sum_j \omega(\hat{\Omega}_j)$ factor, and $\mathcal{Z}_0$ is dominated by $\tau$-independent histories $\bm{\hat{\Omega}}(\tau)=\bm{\hat{\Omega}}(0)$. This leads us to
\begin{equation}\label{eqn:semiclassical_Z}
\begin{split}
    \mathcal{Z}_0 &= \int d\bm{\hat{\Omega}} e^{-\mathcal{S}_0(\bm{\hat{\Omega}})}, \\
    \mathcal{S}_0(\bm{\hat{\Omega}}) &= -\beta J \sum_{\langle jk \rangle} \hat{\Omega}_j \cdot \hat{\Omega}_k + C(\bm{\hat{\Omega}}), 
\end{split}
\end{equation}
where $d\bm{\hat{\Omega}}$ is now an integral over the discrete set of classical spin configurations $\hat{\Omega}_j$ and we have omitted an overall prefactor. 
The term $C(\bm{\hat{\Omega}})$ comes from the singlet projector. This object can be expressed as 
\begin{equation} \label{eqn:f_integral} 
\begin{split}
    C(\bm{\hat{\Omega}}) \equiv - \ln \int dU  \bra{\bm{\hat{\Omega}}} U^{\otimes N}\ket{\bm{\hat{\Omega}}}.
\end{split}
\end{equation}
The theory in Eq.~\eqref{eqn:semiclassical_Z} will suffice for calculations of spin correlations at leading order in $S$, with $S \gg 1$. 

It is not straightforward to obtain a closed-form expression for $C(\bm{\hat{\Omega}})$, which is in principle a rotation- and permutation-symmetric function of the full spin configuration $\bm{\hat{\Omega}}$. However, at large $S$, $C(\bm{\hat{\Omega}})$ takes on a simple approximate form for typical spin configurations. This approximation will be sufficient to understand the entanglement transition that occurs as $\beta$ is varied, and we justify it numerically below. We outline the large $S$ approximation for $C(\bm{\hat{\Omega}})$ here, and provide the technical details in Appendix \ref{app:int_over_U}.

First note that any $\mathrm{SU}(2)$ rotation can be parametrized in the angle-axis representation $U^{\otimes N} = \bigotimes_{j=1}^N e^{i \gamma \hat n\cdot {\vec{S}}_j}$. We can then write 
\begin{equation}\label{eqn:U_overlap}
\begin{split}
    \bra{\bm{\hat{\Omega}}} U^{\otimes N} \ket{\bm{\hat{\Omega}}} = e^{\sum_j \log\braket{\hat{\Omega}_j|R(\hat{n}, \gamma)\hat{\Omega}_j}},
\end{split}
\end{equation}
where $R(\hat n,\gamma)$ is the $3 \times 3$ orthogonal rotation matrix associated with the single-spin unitary $U_j=e^{i\gamma \hat n \cdot \vec{S}_j}$. The magnitudes of the individual overlaps $\braket{\hat{\Omega}_j|R(\hat{n}, \gamma)\hat{\Omega}_j}$ are well-known to decay exponentially on increasing $S$ for fixed $\hat{\Omega}_j$ and $R(\hat{n}, \gamma)$ \cite{Auerbach1994Spins}
\begin{align}
    |\braket{\hat{\Omega}_j|R(\hat{n}, \gamma)\hat{\Omega}_j}| = 2^{-S}\Big( 1 + \hat{\Omega}_j \cdot R(\hat{n}, \gamma)\hat{\Omega}_j\Big)^S.
\end{align}
Therefore, for typical configurations $\bm{\hat{\Omega}}$, the integral in Eq.~\eqref{eqn:f_integral} is dominated by global $\rm{SU}(2)$ operations $U$ with small $\gamma$, specifically $\gamma = \mathcal{O}((SN)^{-1/2})$ at large $S$ and large $N$.

To make analytical progress from here, we neglect contributions from collinear spin configurations. Such contributions are suppressed entropically, and so we expect that this approximation is appropriate above a temperature that goes to zero in the limit of large $S$. The reason for making this approximation is that the behavior of the integral over $\gamma$ is qualitatively different for collinear configurations compared with typical ones: for $\hat{n}$ perfectly aligned with all of the spins the overlap does not decay with $\gamma$ (however, for such $\hat n$, if the collinear spins have a nonzero total magnetization, the integral over $\gamma$ gives zero).


Expanding to second order in $\gamma$ and integrating over this angle, the integral over $U$ in Eq.~\eqref{eqn:f_integral} reduces to an integral over the rotation axis $\hat{n}$. This still takes a rather complicated form:
\begin{equation}\label{eqn:complicatedf}
\begin{split}
    C(\bm{\hat{\Omega}}) &\approx -\ln \int d\hat{n} \, c(\bm{\hat{\Omega}},\hat{n}) + \text{const.} \\
    c(\bm{\hat{\Omega}},\hat{n})&=  e^{-\frac{S}{a[\bm{\Omega}]N}(\hat{n} \cdot \hat{\Omega}_{\rm{tot}})^2}\frac{N a[\bm{\hat{\Omega}}] - 2 S (\hat{n} \cdot \vec{\Omega}_{\rm{tot}})^2}{a[\bm{\hat{\Omega}}]^{5/2}},
\end{split}
\end{equation}
where we have defined
\begin{align}
    \vec{\Omega}_{\rm{tot}} = \sum_j  \hat{\Omega}_j, \quad a[\bm{\hat{\Omega}}] = 1-\frac{1}{N}\sum_j \left(\hat{n} \cdot \hat{\Omega}_j\right)^2.
\end{align} 
An important simplification nevertheless arises in the paramagnetic phase, and also at high temperatures close to the critical point in the ferromagnetic phase (again assuming the phase transition is continuous). To see this, consider a caricature for a spin configuration in the ferromagnetic phase 
\begin{align}
	\hat{\Omega}_j = m \hat{\Omega}_j^{\mathrm{FM}} + \Big[1-\frac{1}{2}m \hat{\Omega}_j^{\mathrm{FM}} \cdot \hat{\xi}_j+\mathcal{O}(m^2)\Big]\hat{\xi}_j,
\end{align}	
where $m$ is the average magnetization, $\hat{\Omega}_j^{\mathrm{FM}}$ is the ordered ferromagnetic configuration, and we assume $m \ll 1$ close to the critical point. The arguments below also apply in the paramagnetic phase by setting $m = 0$. Suppose for simplicity that the three-component unit vectors $\hat{\xi}_j$, representing fluctuations around the ordered state, have random orientations. The spatial average $\overline{\xi_j}$ of the random fluctuations is therefore $\sim N^{-1/2}$ for large $N$ in typical spin configurations. For small $m$ and large $N$ we then have 
\begin{align}
    a[\bm{\hat{\Omega}}] = 1 - \overline{(\hat{n} \cdot \hat{\xi}_j)^2} + \mathcal{O}(m,N^{-1/2}) \approx \frac{2}{3}.\label{eqn:denominatora}
\end{align}
The spatial average $\overline{(\hat{n} \cdot \hat{\xi}_j)^2}$ is approximately an ensemble average over random unit vectors, and therefore approaches a constant that is independent of $\hat{n}$. It is important to stress that this behavior does not depend on detailed statistical properties of $\hat{\Omega}_j$. In particular, it also holds when the fluctuations $\hat{\xi}_j$ are correlated, as they will be in the vicinity of the thermal phase transition. We only require that the different spins $j$ are approximately uniformly distributed over the sphere, up to a small average magnetization. 

Setting $a = 2/3$ in \eqref{eqn:complicatedf} we can carry out the integral over $\hat{n}$ explicitly, and we obtain a simple approximation to the effect of the singlet constraint
\begin{equation}\label{eqn:f_approximate}
\begin{split}
    C(\bm{\hat{\Omega}}) \approx \frac{3}{2}\frac{S}{N}\vec{\Omega}_{\rm{tot}}^2.
\end{split}
\end{equation}
This form is justified for the semiclassical spin configurations $\bm{\hat{\Omega}}$ that are dominant in the path integral when the temperature is above or just below the critical temperature for a continuous phase transition, or equivalently when $m \ll 1$. Moreover we expect that, when symmetrizing a statistical ensemble of spin coherent states (see Appendix~\ref{app:int_over_U} for a general discussion), this approximation to the singlet constraint becomes increasingly accurate as the system size $N$ is increased for \emph{any} value of $S$.

The semiclassical theory for the singlet sector is therefore
\begin{equation} \label{eqn:semiclassical_theory}
    \mathcal{S}_0(\bm{\hat{\Omega}}) = -\beta J\sum_{\langle jk\rangle} \hat{\Omega}_j \cdot \hat{\Omega}_k + \frac{3}{2}\frac{S}{N}\vec{\Omega}_{\text{tot}}^2 + \ldots,
\end{equation}
where the ellipsis denotes contributions that are small when $|\vec{\Omega}_{\rm{tot}}| \ll N$ and $S \gg 1$. 

Using this theory we can compute spin correlations in the singlet sector, with technical details provided in Appendix \ref{app:correlation_function}:
\begin{align}
    \text{Tr}[\rho_0 \vec{S}_i \cdot \vec{S}_j] = S^2\braket{\hat{\Omega}_i \cdot \hat{\Omega}_j}_{\mathcal{S}_0} + \mathcal{O}(S/N).
\label{eqn:semiclassical_corr}
\end{align}
Recall that, through Eqs.~\eqref{eq:distillablelowerbound} and \eqref{eq:negativitylowerbound}, we can relate spin correlations to mixed-state entanglement measures.

\begin{figure}
\centering
\includegraphics[width=1\linewidth]{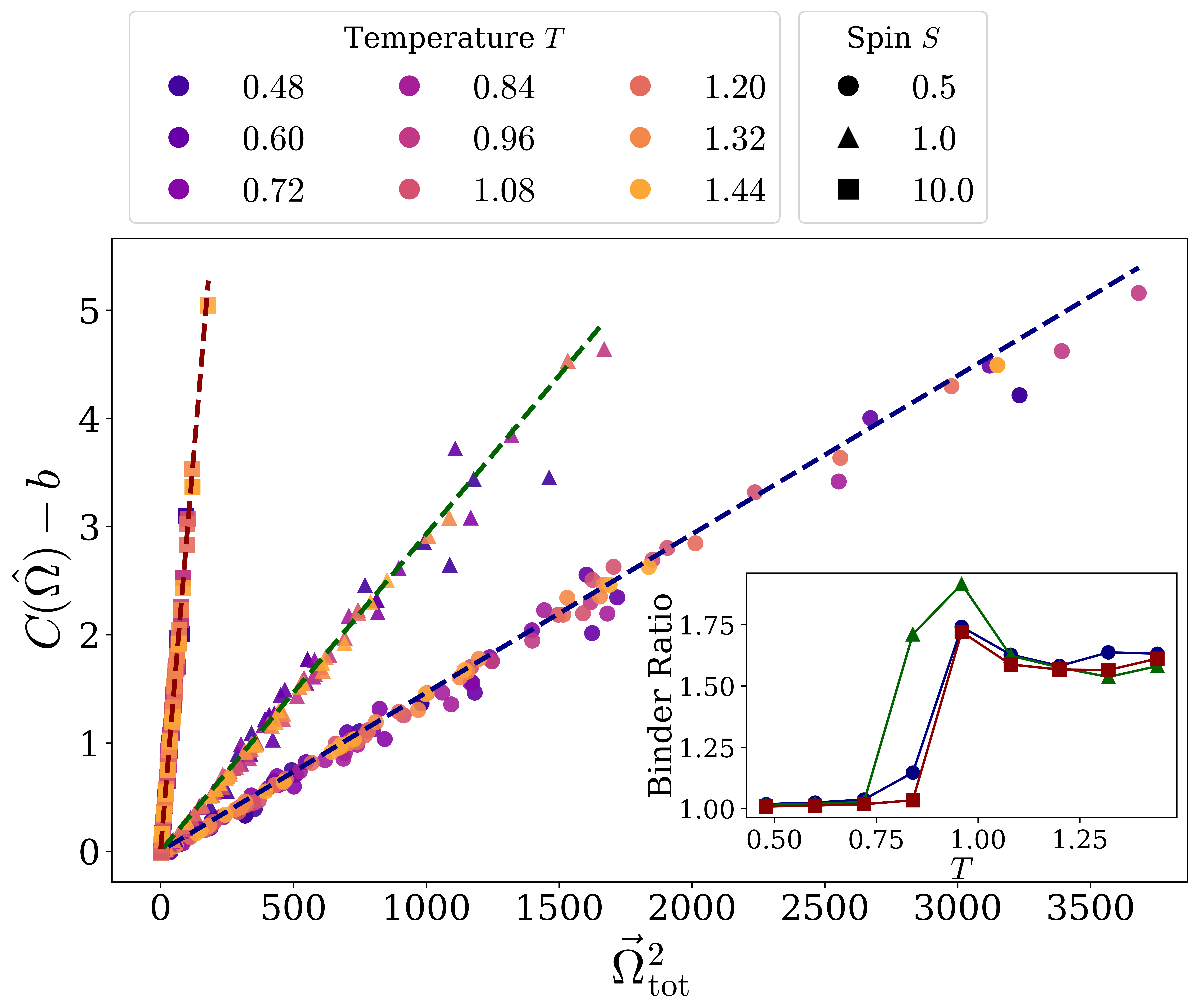}
  \caption{Numerical integration of $C(\bm{\hat{\Omega}})$ (scatter points) shifted to the origin and the analytical prediction $C(\bm{\hat{\Omega}}) = \tfrac{3}{2}\tfrac{S}{N}\vec{\Omega}_{\text{tot}}^2$ (dashed lines), plotted against $\vec{\Omega}_{\text{tot}}^2$, for various $S$ indicated by different shapes. The configurations $\bm{\hat{\Omega}}$ are sampled from the semiclassical theory Eq.~\eqref{eqn:semiclassical_theory} on a three-dimensional cubic lattice with $N = L^3 = 8^3$, and $S = 0.5\, , 1.0\, ,10$, with 15 data points drawn at each temperature. The temperature range is chosen to span both the ordered phase and the disordered phase. The critical temperature separating the two phases is identified by the Binder ratio $B = \langle \vec{\Omega}_{\text{Oct}}^4\rangle/\langle\vec{\Omega}_{\text{Oct}}^2\rangle^2$ where $\vec{\Omega}_{\text{Oct}} = \sum_{x,y,z}^{L/2} \hat{\Omega}_{x,y,z}$ is the total spin living in an octant with linear size $L/2$. As shown in the inset, the temperature range covers both the disordered phase where $B\approx 1.6$, and the ordered phase where $B\approx 1$. For visual clarity, each fixed-$S$ data set has been shifted vertically by $b = \mathrm{avg}\left(\frac{3S}{2N}\vec{\Omega}_{\text{tot}}^2 - C(\bm{\hat{\Omega}})\right)$ with average taken over all data points in this set. Dashed lines are analytical prediction $C(\bm{\hat{\Omega}}) = \frac{3S}{2N}\vec{\Omega}_{\text{tot}}^2$ for different $S$, with blue being $S = 0.5$, green being $S=1.0$, and red being $S = 10$.}
\label{fig:f_integration}
\end{figure}

As a sanity check, note that in the infinite-temperature limit our theory predicts
\begin{equation}
\begin{split}
	\braket{\hat{\Omega}_i \cdot \hat{\Omega}_j}_{\mathcal{S}_0} = -\frac{1}{N} + \ldots,
\end{split}
\end{equation}
at large $S$ and large $N$, where as usual the ellipsis denotes subleading corrections. This agrees with the exact result $\text{Tr}[\rho_0 \vec{S}_i \cdot \vec{S}_j] = -S(S+1)/(N-1)$, which follows from the statement that the total spin squared has expectation value zero. 

We now numerically test our approximation to the singlet constraint. The key question is whether Eq.~\eqref{eqn:f_approximate} is accurate for the kinds of spin configurations that dominate the partition function. To address this we sample spin configurations $\bm{\hat{\Omega}}$ from the semiclassical theory Eq.~\eqref{eqn:semiclassical_theory}, defined for a Heisenberg model on a three-dimensional cubic lattice, and numerically evaluate the integral over $U$ in Eq.~\eqref{eqn:f_integral}.
The result of this integration is exact up to numerical errors, and in Fig.~\ref{fig:f_integration} we compare it with the approximation in Eq.~\eqref{eqn:f_approximate}, finding excellent agreement even for the smallest values of $S$.

This result supports our claim, below Eq.~\eqref{eqn:f_approximate}, that the Gaussian approximation to the singlet constraint is self-consistent and appropriate for all $S$ when $N$ is large. It is, however, important to note that the large-$S$ limit was important for simplifying the path integral representation of the Gibbs state, and for simplifying our expression for the correlation function.

In the inset of Fig.~\ref{fig:f_integration} we also verify that the range of temperatures that we investigate includes both the ordered and the disordered phases of the semiclassical theory. To do this we calculate the Binder ratio $B = \langle \vec{\Omega}_{\text{Oct}}^4\rangle/\langle\vec{\Omega}_{\text{Oct}}^2\rangle^2$, where $\vec{\Omega}_{\text{Oct}} = \sum_{x,y,z}^{L/2} \hat{\Omega}_{x,y,z}$ is the total spin living in an octant with linear size $L/2$, which is expected to approach $1.6$ in the disordered phase and $1$ in the ordered phase. Having arrived at an approximate theory for spin correlations in the singlet sector, which is appropriate in both the low- and high-temperature phases, we now provide evidence that the thermal ordering transition coincides with an entanglement transition. 
  
\subsection{Entanglement transition}\label{subsec:IV.SCT_analysis}

As discussed in Sec.~\ref{sec:entanglement}, the behavior of various mixed-state entanglement measures are related to the half-system total spin. Within the semiclassical theory the expectation value of the half-system total spin is given by
\begin{align}
    \text{Tr}\left[\rho_0 \vec{S}_A^2\right] = S^2\braket{\vec{\Omega}_A^2}_{\mathcal{S}_0} + \mathcal{O}(NS),\label{eqn:halfsys_semiclassical}
\end{align}
where $\vec{\Omega}_A=\sum_{j \in A} \hat \Omega_j$ and the expectation value on the right is computed with respect to the action $\mathcal{S}_0(\bm{\hat{\Omega}})$. 

Here we investigate the statistical properties of the half-system total spin and therefore the scaling of various entanglement measures with $N$ in different phases of the semiclassical theory, and the transition between them. We discuss the disordered and ordered phases in Secs.~\ref{sec:disordered} and Sec.~\ref{sec:ordered}, respectively.

First recall that, in the vicinity of the finite-temperature ordering transition in the standard Gibbs state, we can understand correlations in terms of a coarse-grained three-component vector field $\vec{\Omega}(\vec{x})$. We define $\vec{\Omega}(\vec{x})$ as the \emph{sum} of the unit vectors $\hat \Omega_j$ within a cubic subregion of linear extent $\ell$, with $\ell$ much larger than the lattice spacing but much smaller than the system linear extent $L$. In terms of this field the partition function can be described by 
\begin{equation}\label{eqn:standard_theory}
\begin{split}
    \mathcal{Z} &\propto \int D \vec{\Omega}(\vec{x}) e^{-\mathcal{S}(\vec{\Omega})} \\
    \mathcal{S}(\vec{\Omega}) &= \int d^d \vec{x}\Big[ r\partial_j \Omega^a \partial_j \Omega^a + t \Omega^a \Omega^a + u \Omega^a \Omega^a \Omega^b \Omega^b + \ldots\Big]
\end{split}
\end{equation}
where here the integral in $\mathcal{Z}$ is over fields $\vec{\Omega}(\vec{x})$ in $d$ spatial dimensions, and our convention is that repeated vector indices $a=x,y,z$ are summed. The phenomenological parameter $t$ is the standard reduced temperature, $r$ is associated with the spin wave stiffness in the ordered phase, and $u$ constrains the magnitudes of spin fluctuations.

We work in a total volume that we denote $N=L^d$ and with periodic boundary conditions. The theory is symmetric under global rotations in spin space ${\Omega}^a \to R^{ab} {\Omega}^b$. For $d \geq 3$ this theory exhibits a phase transition in the classical $\rm{O}(3)$ universality class tuned by the parameter $t$. Although the lattice Hamiltonian explicitly breaks rotational symmetry in real space, this is restored in the coarse-grained theory since the breaking of this symmetry is irrelevant; these irrelevant contributions and others are represented by the ellipsis in Eq.~\eqref{eqn:standard_theory}.

In general the singlet constraint enters the partition function as a complicated and nonlocal function of the full spin configuration. However, as we have shown above, provided the global magnetization is well below its saturation value, such as in the vicinity of the critical point, this constraint can be approximated by a coarse-grained property of the spin configuration. Defining the inverse Fourier transform $\Omega^a(\vec{x}) = L^{-d/2} \sum_{\vec{k}} \tilde{\Omega}^a(\vec{k}) e^{-i \vec{k}\cdot \vec{x}}$,
where the sum runs over $d$-component wavevectors $\vec{k}$ with $k_j = (2\pi /L) \times \mathbb{Z}$, we have
\begin{equation} \label{eq:S0}
\begin{split}
    &\mathcal{S}_0(\vec{\Omega}) = \sum_{\vec{k}}[t + r\vec{k}^2] \tilde \Omega^a(\vec{k}) \tilde \Omega^a(-\vec{k})  + bS \tilde \Omega^a(0) \tilde \Omega^a(0)\\
	&+ u L^{-d} \sum_{\vec{k}_1+\vec{k}_2+\vec{k}_3+\vec{k}_4=0}\tilde \Omega^a(\vec{k}_1)\tilde\Omega^a(\vec{k}_2)\tilde\Omega^b(\vec{k}_3)\tilde\Omega^b(\vec{k}_4) + \ldots, 
\end{split}
\end{equation}
where $b$ is a constant of order unity; setting $b=0$ recovers the unconstrained theory $\mathcal{S}(\vec{\Omega})$. 

The nature of the ordering transition in ferromagnets is conventionally understood through the renormalization group. In our coarse-grained theory the singlet constraint appears only as a modified `mass' term  only for the $\vec{k}=0$ mode; for this reason the constraint does not affect the momentum-space renormalization group transformations at any finite order. The implication is that, if there is an ordering transition on varying $t$ in $\mathcal{S}(\vec{\Omega})$, there is a transition in $\mathcal{S}_0(\vec{\Omega})$ and therefore the singlet Gibbs state. 

Within this theory we can now analyze the distillable entanglement and the logarithmic entanglement negativity. 
Recall the lower bounds of these quantities from Eqs.~\eqref{eq:distillablelowerbound0} and \eqref{eq:negativitylowerbound0},
\begin{equation}\label{eq:EDENlargeS}
\begin{split}
        E^{D\;<}_{A:B} &= \sum_s p(s)\log_2 (2s+1), \\
    E^{N\;<}_{A:B} &= \log_2 \sum_s p(s) (2s+1),
\end{split}
\end{equation}
where $p(s)=\sum_{ab}p(s,a,b)$ with $p(s,a,b)$ defined in Eqn.~\eqref{eq:p_sab}, and the $\sum_s$ above runs over $s \geq 0$. 

To determine the form of the probability distribution $p(s)$ we consider $\text{Tr}[\rho_0 \vec{S}_A^{2k}]$, with $k$ being a positive integer, which can be expressed as the expectation value of $[s(s+1)]^k$ with respect to $p(s)$. Alternatively, we can compute $\text{Tr}[\rho_0 \vec{S}_A^{2k}]$ within the semiclassical theory; see Appendix~\ref{app:correlation_function} for details. The result is $S^{2k}\braket{|\vec{\Omega}_A|^{2k}}_{\mathcal{S}_0}$ to leading order in $S$. Therefore, 
\begin{align}
    \sum_s p(s) [s(s+1)]^k =S^{2k}\braket{|\vec{\Omega}_A|^{2k}}_{\mathcal{S}_0} + \ldots,\label{eq:SA2ksemiclassics}
\end{align}
where $\vec{\Omega}_A = \int_A d^dx\, \vec{\Omega}(\vec{x})$ and the ellipses denote corrections that are subleading for $S,\, N \gg 1$. At leading order in $S$ the simplest possibility is that the distribution of $s$ is exactly the distribution of $S|\vec{\Omega}_A|$. This assumption ensures that all of the moments in Eq.~\eqref{eq:SA2ksemiclassics} match, and is the basis for our semiclassical calculation of mixed-state entanglement. Below we analyze the scaling of these expressions in each phase.

\subsubsection{Disordered Phase}\label{sec:disordered}

In the disordered phase the quartic term is irrelevant and the theory is effectively Gaussian in $\vec{\Omega}(\vec{x})$:
\begin{equation}
\begin{split}
    \mathcal{S}_0(\vec{\Omega}) &\approx \int d^d \vec{x}\left( r\partial_j \Omega^a(\vec{x}) \partial_j \Omega^a(\vec{x}) + t \Omega^a(\vec{x}) \Omega^a(\vec{x})\right)\\
    & + b\frac{S}{L^d}\int d^d \vec{x}_1 d^d \vec{x}_2 \Omega^a (\vec{x}_1)\Omega^a (\vec{x}_2).
\end{split}
\end{equation}
Within this theory it is straightforward to evaluate the moments $\text{Tr}[\rho_0 \vec{S}_A^{2k}]$ using Eq.~\eqref{eq:SA2ksemiclassics}. From Wick's theorem we find \begin{align}
    \text{Tr}[\rho_0 \vec{S}_A^{2k}]\approx S^{2k}\langle |\vec{\Omega}_A|^{2k}\rangle_{\mathcal{S}_0} \approx c_k (g S^{2}L^d \xi^d)^k,
\label{eq:disorderedWick}
\end{align}
up to terms that are sub-leading in $S$ and $L$. Here $c_k$ is a combinatorial factor that comes from pairing vector indices and $g$ is a geometric factor that depends on the shape of $A$. For sufficiently large $t>0$ that the Ginzburg criterion is satisfied, the correlation length $\xi \sim (t/r)^{-1/2}$. To match these moments with those obtained by averaging over $p(s)$ through
\begin{align}
    \text{Tr}[\rho_0 \vec{S}_A^{2k}] = \sum_s p(s) [s(s+1)]^k,
\end{align} 
the distribution $p(s)$ needs to be
\begin{align}
    p(s) \propto s^2 e^{-s^2/[2(g S^2 L^d \xi^d)]}.
    \label{eq:ps}
\end{align}
Another way to understand this distribution is that the moments \eqref{eq:disorderedWick} originate from a Gaussian distribution of $\vec{\Omega}_A$, which upon angular integration leads to the distribution \eqref{eq:ps} of $|\vec{\Omega}_A|\sim s$. 

Using the above distribution $p(s)$ to compute the lower bounds on the distillable entanglement and entanglement negativity gives
\begin{equation}\label{eqn:disorder_scaling}
\begin{split}
    E^{D\;<}_{A:B} &= \frac{d}{2}\big(\log_2 L  + \log_2\xi\big) + \log_2 S + \ldots,\\
    E^{N\;<}_{A:B} &= \frac{d}{2}\big(\log_2 L  + \log_2\xi\big) + \log_2 S + \ldots, 
\end{split}
\end{equation}
up to corrections of order unity. 

These lower bounds are appropriate in the regime where $\xi \ll L$. For large $L$ and finite $\xi$ it is clear that we recover the infinite-temperature  result $\frac{d}{2}\log_2 L$. Interestingly, as the temperature is lowered in a disordered phase, and the correlation length diverges, our theory indicates that the distillable entanglement and logarithmic negativity increase. In one spatial dimension we will see that both approach $d \log_2 L$, up to subleading corrections, and we confirm this behavior numerically in Sec.~\ref{sec:numerics}. In dimensions higher than one, as we discuss in the following sections, subtleties arise that depend on system geometry. There is nevertheless an entanglement transition at the thermal phase transition in the singlet sector.

\subsubsection{Ordered phase}\label{sec:ordered}

The exact singlet sector constraint forbids any spatially uniform configuration: a state with fully aligned spins carries maximal total spin, and thus has zero support in the singlet sector. Crucially, however, the constraint only penalizes the formation of global magnetization, but does not forbid \emph{local} alignment of the spins. As a consequence, the scaling of subsystem spin with volume can still undergo a transition as the temperature is lowered in spatial dimension $d \geq 3$.  

To determine the statistical properties of the sub-system spin, we study the fluctuations about an ordered non-uniform saddle point of $\mathcal{S}_0(\vec{\Omega})$ on a cubic lattice with periodic boundary conditions:
\begin{align}
    \bar{\Omega}^a(\vec{x}) = \sum_{i=1}^d m^a_i\cos\big[K_i x_i + \phi^a_i\big]. \label{eq:saddle}
\end{align}
Here $\bar{\Omega}^a(\vec{x})$ with $a=x,y,z$ are three components of the vector field $\bar{\Omega}(\vec{x})$, $m^a_i$ is the amplitude of an oscillation in the $a$ component of the spins along the $i$ direction in space, $K_i = 2\pi/L_i$ is the minimum nonzero wavenumber in the $i$ direction, which has linear extent $L_i$, and $\phi^a_i$ parametrize translations of the saddle point in real space. Since $\bar{\Omega}(\vec{x})$ has no weight at wavevector $\vec{k}=0$, it is not penalized by the singlet constraint in $\mathcal{S}_0(\vec{\Omega})$. 

Inserting Eq.~\eqref{eq:saddle} into $\mathcal{S}_0(\vec{\Omega})$ it can be verified that the condition for $\bar{\Omega}(\vec{x})$ to be a saddle point is
\begin{align}
    t + 2u |\bar{\Omega}(\vec{x})|^2 + r\vec{K}^2 = 0, \label{eq:spcondition}
\end{align}
which implies that $|\bar{\Omega}(\vec{x})|$ is spatially uniform. The solution is to only allow $m^a_i \neq 0$ for one of the spatial directions $i=1,\ldots,d$, to choose the phases $\phi^a_i$ for this $i$ such that $|\bar{\Omega}(\vec{x})|$ is uniform, and to impose
\begin{equation}\label{eq:spsolution}
\begin{split}
        \sum_a (m^a_i)^2 \cos[2\phi^a_i] &= \sum_a (m^a_i)^2 \sin[2\phi^a_i] = 0,\\
    \sum_a (m^a_i)^2 &= -\frac{r\vec{K}^2 + t}{u}.
\end{split}
\end{equation}
This implies that $m^a_i \neq 0$ for $t < -r|\vec{K}|^2$, which we identify as the ordered phase. Note that the shift in the critical temperature, relative to that in the standard Gibbs state, vanishes at large $L$.

With all the aforementioned conditions, from here on we focus on the saddle point of the form
\begin{align}\label{eqn:saddle_ansatz}
    \bar{\Omega}(\vec{x}) = m(\cos[\vec{K}\cdot \vec{x}],\, \sin[\vec{K}\cdot \vec{x}],\, 0),
\end{align}
where $\vec{K}$ is aligned with coordinate axis $i$, i.e. $K_i=2\pi/L_i$ and $K_{j \neq i}=0$. Note that this saddle point is one member of a continuous family related by rotations in spin space $\Omega^a(\vec{x}) \to R^{ab} \Omega^b(\vec{x})$ and, in a translation-invariant system, shifts of the phase $\phi^a_i \to \phi^a_i + \delta \phi^a_i$. In a cubic geometry with all $L_i$ equal there are $d$ distinct families of saddle points related by $\pi/2$ spatial rotations.
On the other hand, for systems with fixed aspect ratios $L_j/L_i$, in large systems the partition function is dominated by the family of saddle points with $\vec{K}$ along the coordinate axis with the largest $L_i$. In Appendix~\ref{app:Hessian} we show that the configurations in Eq.~\eqref{eqn:saddle_ansatz} are locally stable minima of the free energy in the ordered phase.

Let us first consider the statistical properties of the subsystem total spin in a system with a distinct maximum $L_i$ which we choose to be along the $i=x$ direction. Following this it will be straightforward to obtain the result for a cubic geometry (with unit aspect ratios). In the following it will often be convenient to write our results in terms of $L = (\prod_i L_i)^{1/d}$.

\begin{figure}
\includegraphics[width=0.9\linewidth]{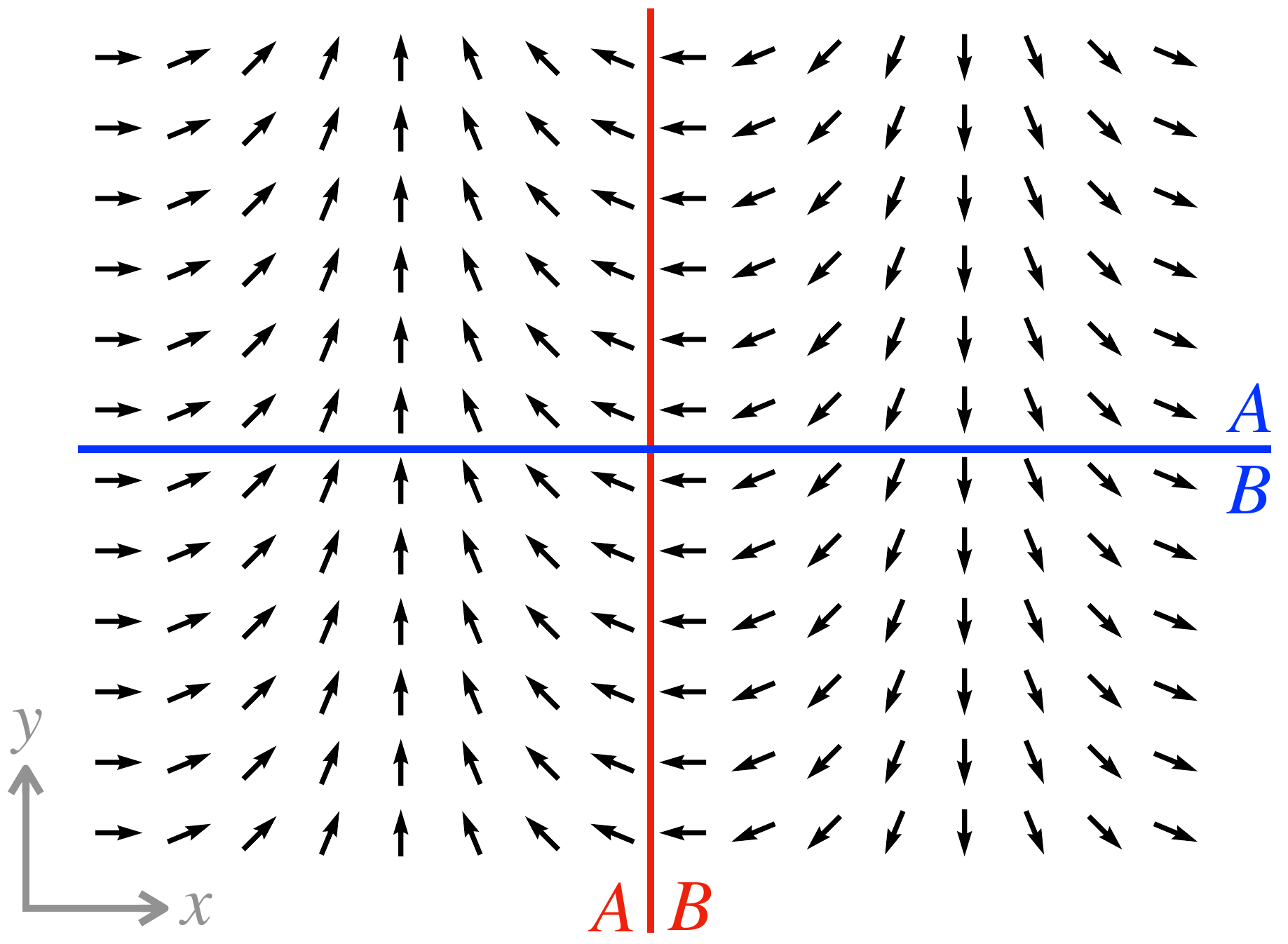}
\caption{Two-dimensional spatial slice through a $d$-dimensional saddle point configuration $\bar{\Omega}_j$ (black arrows) in an anisotropic system with $L_x > L_{i \neq x}$. These configurations are only stable against thermal fluctuations in $d \geq 3$ and at low temperatures. The wavevector of the saddle point is along the spatial $x$ direction. For the bipartition $AB$ indicated by the red line the subsystem total spin $\vec{\Omega}_A$ is dominated by the saddle point. In this case we find the lower bounds on mixed-state entanglement measures in Eq.~\eqref{eq:twist_perp_cut}. For the blue bipartition the subsystem total spin for the saddle point configuration vanishes $\bar{\Omega}_A=0$, and the statistical properties of $\vec{\Omega}_A$ are instead controlled by long-wavelength twists of the spin configuration. The corresponding lower bounds on mixed-state entanglement are given in Eq.~\eqref{eq:twist_para_cut}.}
\label{fig:bipartitions}
\end{figure}

If the entanglement cut separating $A$ from $B$ is normal to the $\hat{x}$ direction, the subsystem total spin at the saddle point $\bar{\Omega}_A = \int_A d^d x\,\bar{\Omega}(\vec{x})$ has magnitude $|\bar{\Omega}_A| = (m/\pi) L^d$. This situation is illustrated as the red vertical bipartition in Fig.~\ref{fig:bipartitions}. There are fluctuations around the saddle-point value, but these lead to changes in $|\vec{\Omega}_A|$ that are parametrically smaller in system size than $|\bar{\Omega}_A|$. This implies that the distribution of $p(s)$ is sharply peaked at $\sim mS L^d$, with width parametrically smaller in $L$. We therefore find that at leading order in $L$,
\begin{equation}\label{eq:twist_perp_cut}
\begin{split}
    E^{D\;<}_{A:B} &= d \, \log_2 L + \log_2 mS + \ldots,\\
     E^{N\;<}_{A:B} &= d \, \log_2 L + \log_2 mS + \ldots,
\end{split}
\end{equation}
and we note that $m$ increases as the temperature $T$ is decreased. In particular we expect that $m \to 1$ as $T \to 0$.

On the other hand, if the $\hat{x}$ direction is in the plane of the entanglement cut, as illustrated by the blue horizontal bipartition in Fig.~\ref{fig:bipartitions}, the contribution to $\vec{\Omega}_A$ from the saddle point vanishes, $\bar{\Omega}_A=0$, and it is necessary to account for fluctuations in $\vec{\Omega}_A$ around this value. In the following we consider an entanglement cut normal to the $\hat{y}$ direction, as illustrated in Fig.~\ref{fig:bipartitions}, although the discussion below works for any one of the $d-1$ axis directions orthogonal to $\hat{x}$.

We perform a systematic analysis of the different kinds of fluctuations around the saddle point in Appendix~\ref{app:Hessian}, and here we focus on the ones that are intuitive and control the scaling of $\vec{\Omega}_A$ with $L$. These are Goldstone fluctuations, i.e. smoothly varying rotations of the saddle-point spin configuration $\bar{\Omega}(\vec{x})$. Moreover, it will suffice to focus on rotations about an axis in the plane of the saddle point $\bar{\Omega}(\vec{x})$, and a rotation angle that varies in space as $\theta(\vec{x}) = \theta \cos(\vec{K}\cdot \vec{x}) \sin(\vec{k}\cdot \vec{x})$ where $\vec{k}$ is normal to both $\vec{K}$ and to the entanglement cut, in this case along the spatial $y$ direction ${\vec{k} = (2\pi n/L_y)\hat y}$.

At small $\theta$ the change in subsystem total spin due to such a fluctuation scales as
\begin{align}
    |\vec{\Omega}_A| \sim 
    \begin{cases}
     L^d m\,\theta/n \quad &n \;\rm{odd}\\ 
    0 \quad &n\;\rm{even}.
    \end{cases}
\label{eqn:OmegaAGoldstone}
\end{align}
Meanwhile, expanding the semiclassical theory to quadratic order in the small fluctuation $\theta$ gives a change
\begin{align}
    \delta \mathcal{S}_0 \sim L^d r |\vec{k}|^2 (m\theta)^2 \sim r L^{-(d+2)}n^4 |\vec{\Omega}_A|^2 \label{eqn:goldstone_deltaS}
\end{align}
in the action.

Following the same idea as in the disordered phase where the distribution of  $|\vec{\Omega}_A|$ is a Gaussian, here by matching the moments of $\vec{S}_A^2$ in the singlet ensemble and the moments of $\vec{\Omega}_A^2$ in the quadratic fluctuation theory Eqn.~\eqref{eqn:goldstone_deltaS}, we find the distribution of $s$ to be
\begin{align}
    p(s) \propto s^2 e^{-s^2/[2(g' r^{-1}S^2 L^{d+2})]},
    \label{eq:ps_ordered}
\end{align}
where $g'$ is a geometric factor of order unity. Using this distribution and Eqs.~\eqref{eq:EDENlargeS} we find 
\begin{equation}\label{eq:twist_para_cut}
\begin{split}
    E^{D\;<}_{A:B} &= \big(1 + d/2\big) \, \log_2 L + \log_2 (S/\sqrt{r}) + \ldots,\\
    E^{N\;<}_{A:B} &= \big(1 + d/2\big) \, \log_2 L + \log_2 (S/\sqrt{r}) + \ldots,
\end{split}
\end{equation}
in anisotropic systems that have a unique maximum $L_i$, and for entanglement cuts containing the corresponding axis. 

Note that the scaling with $L$ in Eq.~\eqref{eq:twist_para_cut} is distinct from both the paramagnetic phase and from the case where the entanglement cut is normal to $\vec{K}$ in the ordered phase. Additionally, in a concrete lattice model we expect that the parameter $r$ will increase as the temperature is reduced, corresponding to a suppression of fluctuations in subsystem total spin. So we expect the subleading corrections (here $\log_2 (S/\sqrt{r})$) to \emph{decrease} as the temperature is decreased. 

More care is required in an isotropic system for several reasons. First, when all $L_i$ are equal, there are $d$ distinct families of saddle points corresponding to $\vec{K}$ along the $d$ different coordinate axes. The full distribution of $|\vec{\Omega}_A|$ is therefore a mixture of the distributions where $\vec{K}$ is and is not normal to the entanglement cut, with relative weights $1/d$ and $(d-1)/d$, respectively. While the contribution with $\vec{K}$ normal to the entanglement cut is quite simple, being concentrated around $|\vec{\Omega}_A| \sim L^d$ as above, contributions with $\vec{K}$ in the plane of the entanglement cut are subtle. In particular, a curious feature in the isotropic geometry is that the free-energy cost of one of the low-energy spin fluctuations vanishes at quadratic order. We discuss this in detail in Appendix~\ref{app:Hessian}, but the result is that this spurious soft mode acquires a free energy cost at quartic order, and so with $\vec{K}$ in the plane of the entanglement cut we still find $|\vec{\Omega}_A| \sim L^{1+d/2}$ as in the anisotropic case.

For all $L_i$ equal, averaging over contributions from saddle points with $\vec{K}$ along the $d$ different directions leads to 
\begin{equation}
\begin{split}
    \langle \log_2 |\vec{\Omega}_A|\rangle_{\mathcal{S}_0} &= \Big(\frac{d}{2}+\frac{3}{2}-\frac{1}{d}\Big)\log_2 L + \ldots,\\
    \log_2 \langle |\vec{\Omega}_A|\rangle_{\mathcal{S}_0} &= d \log L + \ldots.
\end{split}
\end{equation}
In each of these expressions the ellipsis represents terms that are subleading. The logarithm of the average $|\vec{\Omega}_A|$ is therefore parametrically larger than the average of the logarithm. The resulting lower bounds on distillable entanglement and logarithmic entanglement negativity are then
\begin{equation}\label{eq:isotropic_ordered}
\begin{split}
    E^{D\;<}_{A:B} &\approx \Big(\frac{d}{2}+\frac{3}{2}-\frac{1}{d}\Big)\log_2 L \, + \log_2 S \, + f_D(T),\\
    E^{N\;<}_{A:B} &\approx d \log_2 L + \log_2 S + f_N(T),
\end{split}
\end{equation}
where $f_D(T)$ and $f_N(T)$ are temperature-dependent functions that depend on the phenomenological parameters in our theory.

Equations~\eqref{eq:twist_perp_cut} and \eqref{eq:twist_para_cut}, as well as their analogs \eqref{eq:isotropic_ordered} in isotropic systems are some of our central results. On decreasing the temperature in a large but finite-size system, and passing through the (finite-size analog) of the thermal phase transition, our lower bounds on mixed-state entanglement experience discontinuous jumps that are parametrically large in the system size.

\subsection{Monte Carlo}\label{subsec:monte_carlo}

The basic premise of our semiclassical theory for spin correlations is that, at large $S$, the statistical properties of the eigenvalue $s(s+1)$ of $\vec{S}_A^2$ in the quantum state $\rho_0$ match the statistics of the object $(S|\vec{\Omega}_A|)^2$ computed with respect to the action $\mathcal{S}_0(\bm{\hat{\Omega}})$. 

To support our analysis above, in this section we use classical Monte Carlo simulations of the theory in Eq.~\eqref{eqn:semiclassical_theory} (neglecting terms that are subleading at large $S$) to numerically compute properties of $|\vec{\Omega}_A|$ that are associated with the distillable entanglement and the logarithmic entanglement negativity. The lower bound on distillable entanglement is given by the average of $\log_2(2s+1)$ with respect to $p(s)$ [see Eq.~\eqref{eq:distillablelowerbound0}], and is therefore associated with the average of $\log_2 |\vec{\Omega}_A|$ in the semiclassical theory, up to corrections that are sub-leading in $L$. To extract the scaling of this quantity with $L$ we will compute the average ${\langle\log_2 |\vec{\Omega}_A|\rangle_{\mathcal{S}_0}}$. The lower bound on logarithmic negativity is instead associated with the logarithm of the average of $2s+1$, so to extract its scaling with $L$ in Monte Carlo we compute $\log_2\langle |\vec{\Omega}_A|\rangle_{\mathcal{S}_0}$. 

Our approach will be to average $|\vec{\Omega}_A|$ and its logarithm over Monte Carlo runs in which the unit vectors $\hat \Omega_j$ are updated according to the statistical weight $e^{-\mathcal{S}_0(\bm{\hat \Omega})}$ with $\mathcal{S}_0$ given by  Eq.~\eqref{eqn:semiclassical_theory}. Since we are primarily interested in the entanglement transition, which coincides with the thermal phase transition in our semiclassical theory, we focus on spin systems on cubic lattices in $d=3$ and fix $S=50$. We consider both isotropic and anisotropic geometries, and in the latter case we set $L_x > L_y=L_z$ and the bipartition of the system into $A$ and $B$ to be across a spatial $y-z$ plane. 
For each system size and geometry we perform $50$ independent annealing runs.

\begin{figure}
\includegraphics[width=\linewidth]{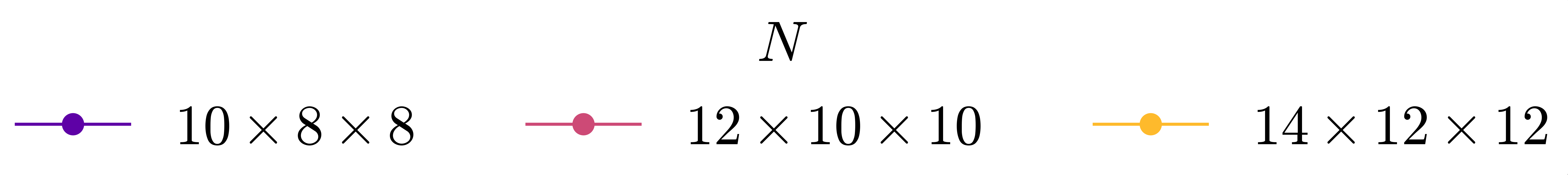}
\includegraphics[width=\linewidth]{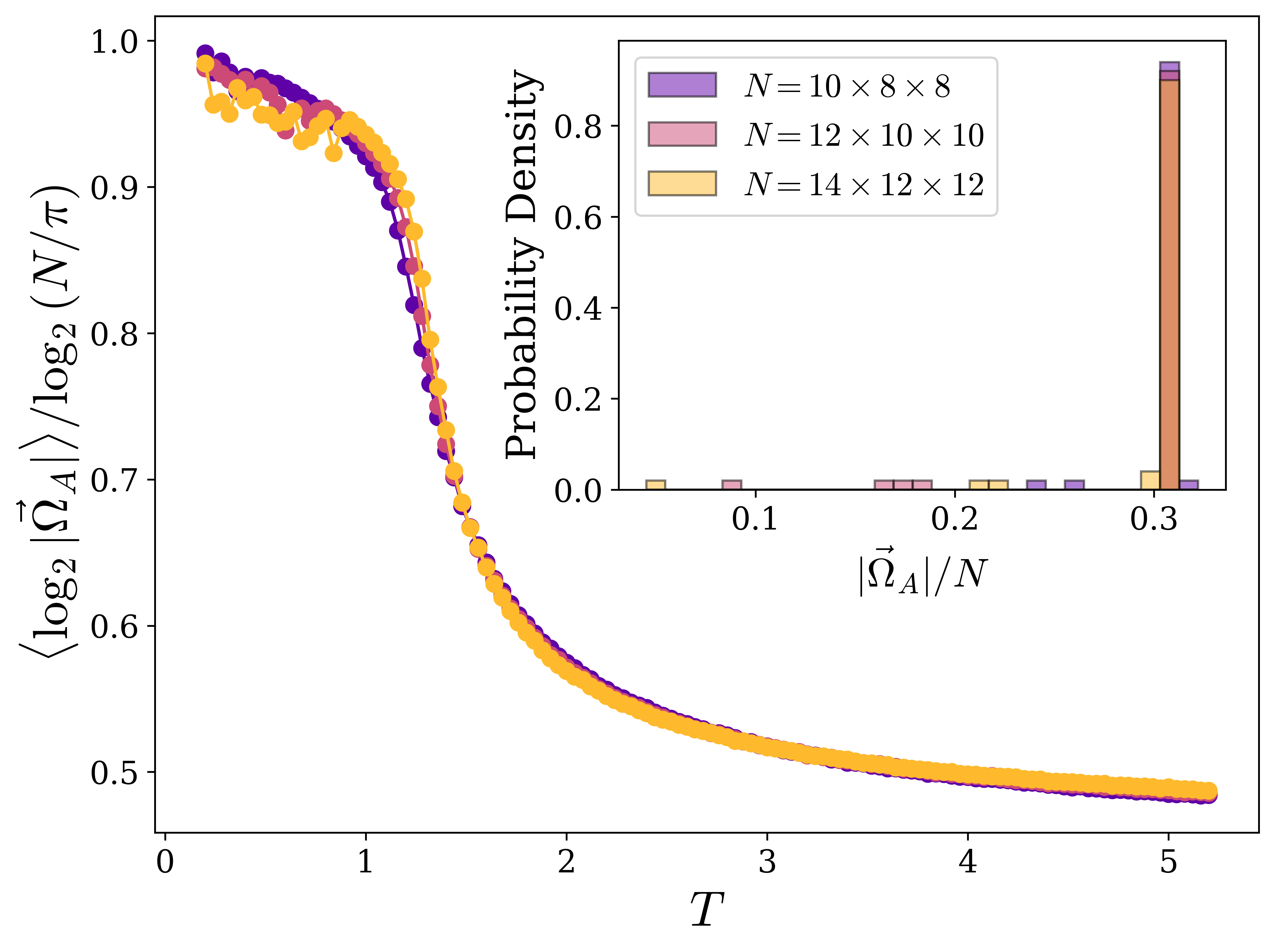} 
\includegraphics[width=\linewidth]{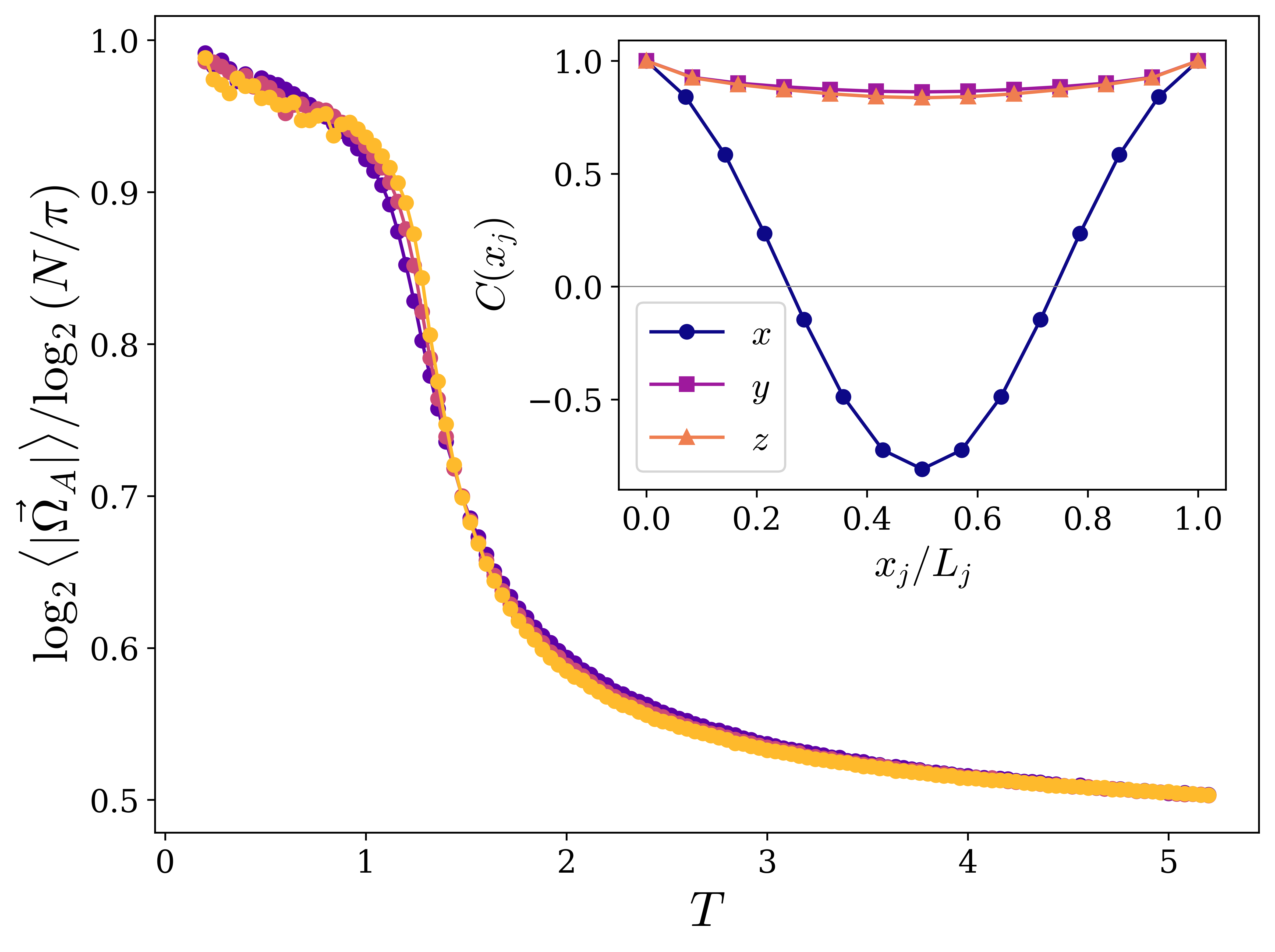} 
\caption{Monte Carlo calculations of the lower bounds on (a) distillable entanglement and (b) logarithmic negativity in anisotropic systems with $S=50$. For $S\gg 1$ the full lower bounds are approximately $\langle \log_2 |\vec{\Omega}_A|\rangle$ and  $\log_2\langle |\vec{\Omega}_A|\rangle$, respectively. Here the system dimensions are $N =L_x \times L_y \times L_z$ (see legend), where $L_x > L_y=L_z$, and the bipartition of the system into equal halves $A$ and $B$ is across a spatial $y-z$ plane. In the main panels we find both $\langle \log_2 |\vec{\Omega}_A|\rangle/\log_2(N/\pi)$ and $ \log_2 \langle|\vec{\Omega}_A|\rangle/\log_2(N/\pi)$ to approach 1 at low temperatures as predicted in Eq.~\eqref{eq:twist_perp_cut}, and to approach 0.5 at high temperatures as predicted in Eq.~\eqref{eqn:disorder_scaling}, where the $\log_2(\pi)$ is a subleading (in $L$) factor that should be negligible in thermodynamic limit. In the inset of (a) we show the histogram of $|\vec{\Omega}_A|/N$ computed at $T=0.2$ for the 50 independent annealing runs for each system size, with the half-system cut in the $y-z$ plane. In the inset of (b) we show the spin-spin correlation function $C(x_j) = {\langle \hat{\Omega}(\vec{x}) \cdot \hat{\Omega}(\vec{x}+x_j \hat{e}_j)\rangle_{\mathcal{S}_0}}$, computed at $T=0.2$ and $L=12$, along the three different lattice directions $\hat{e}_{j=x,y,z}$.}
\label{fig:ED_anisotropic}
\end{figure}

To test our theory we first investigate anisotropic systems. When the half-system cut is in the $y-z$ plane, we expect both $\langle\log_2|\vec{\Omega}_A|\rangle$ and $\log_2\langle|\vec{\Omega}_A|\rangle$ to approach $\log_2(N/\pi)$ at low temperatures [Eq.~\eqref{eq:twist_perp_cut}], and to approach $\frac{1}{2}\log_2(N/\pi)$ at high temperatures [Eq.~\eqref{eqn:disorder_scaling}]. The $\log_2(\pi)$ offset arises from integrating the saddle point configuration $\bar{\Omega}(x)$ over space, and is subleading in the thermodynamic limit. We confirm this behavior in the main panels of Figs.~\ref{fig:ED_anisotropic}(a) and \ref{fig:ED_anisotropic}(b) respectively.
 
In the inset of Fig.~\ref{fig:ED_anisotropic}(a) we also show that $|\vec{\Omega}_A|/N$ is concentrated around its saddle-point value $1/\pi \approx 0.32$ in the low-temperature phase at $T=0.2$, as expected. The inset of Fig.~\ref{fig:ED_anisotropic}(b) shows that the low-temperature spin configuration has a distinct sinusoidal twist along the $\hat{x}$ direction, consistent with our picture above.

\begin{figure}
  \includegraphics[width=0.7\linewidth]{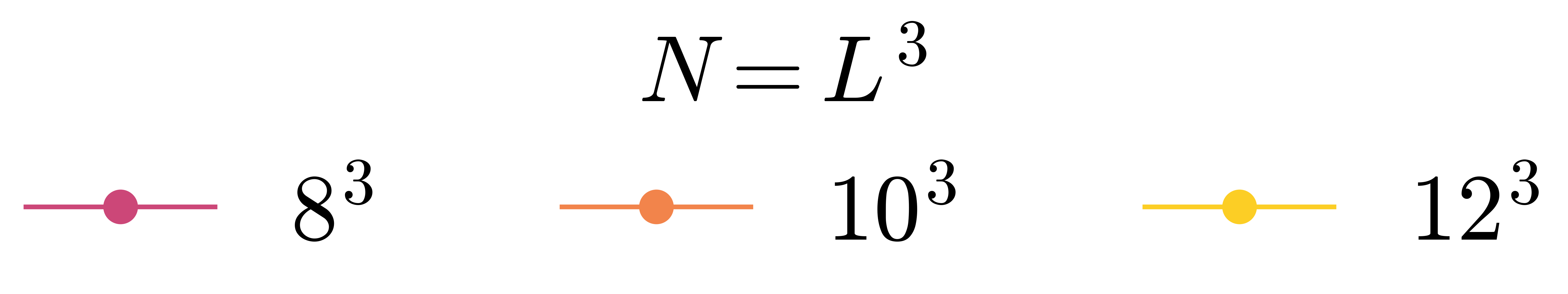}
  \includegraphics[width=\linewidth]{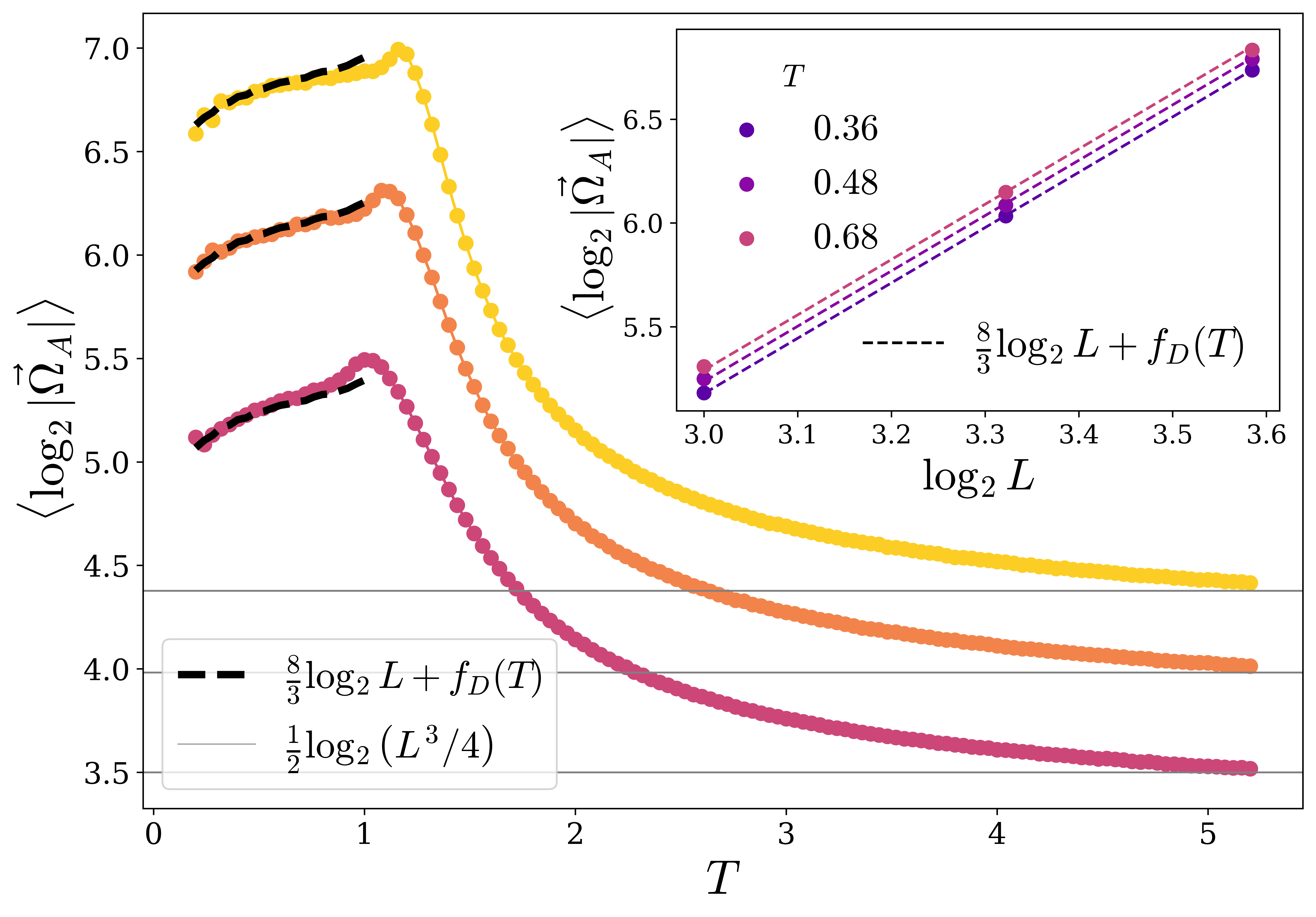} 
  \includegraphics[width=\linewidth]{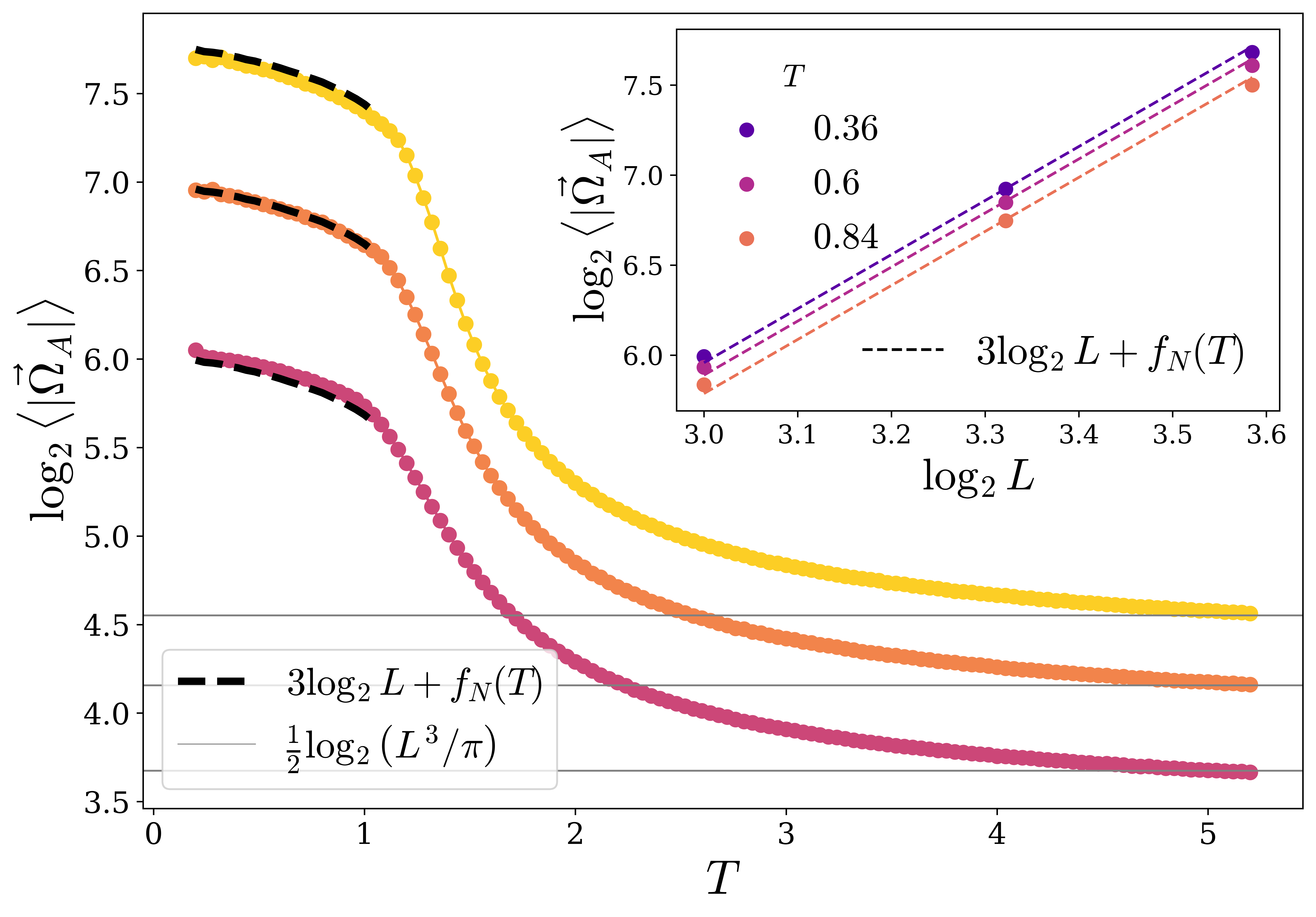}
\caption{Monte Carlo calculations of (a) $\langle \log_2 |\vec{\Omega}_A|\rangle$ and (b) $\log_2 \langle|\vec{\Omega}_A|\rangle$ in isotropic systems $L = L_{x,y,z}$ with $L=8,10,12$ (legend) and $S=50$. Here we average these quantities over equal bipartitions $AB$ of the system across planes normal to each of the three coordinate axes. In the inset of (a) we show $\langle \log_2 |\vec{\Omega}_A|\rangle$ versus the prediction $(8/3)\log_2 L + f_D(T)$ in Eq.~\eqref{eq:isotropic_ordered}, where at each $T$ the object $f_D(T)$ is determined via a single-parameter fit to $\langle \log_2 |\vec{\Omega}_A|\rangle$ versus $\log_2 L$ (see inset). The dashed black lines in the main panel of (a) show $(8/3)\log_2 L + f_D(T)$ for each $L$. In the main panel of (b), dashed black lines show $3\log_2 L + f_N(T)$, where at each $T$ the object $f_N(T)$ is again a fitting parameter. Thin horizontal lines in both (a) and (b) show the expected behavior $\frac{1}{2}\log_2 (L^3/4)$ and $\frac{1}{2}\log_2 (L^3/\pi)$, respectively,  at ${T \to \infty}$ for each $L$.}
\label{fig:EDisotropic}
\end{figure}

For a cubic geometry (with $L_x=L_y=L_z$), the Monte Carlo calculations of $\langle\log_2|\vec{\Omega}_A|\rangle$ and $\log_2\langle |\vec{\Omega}_A|\rangle$ are shown in Figs.~\ref{fig:EDisotropic}(a) and \ref{fig:EDisotropic}(b), respectively, where $\langle \cdots \rangle$ indicates an average over the ensemble of spin configurations as well as over symmetry-related bipartitions. At high temperatures we find that these quantities approach $\frac{1}{2}\log_2(L^3/4)$ and $\frac{1}{2}\log_2(L^3/\pi)$, respectively, as predicted in Eq.~\eqref{eqn:disorder_scaling}, where the constant contributions $\log_2(4)$ and $\log_2(\pi)$ have non-negligible effects on the results in finite-size systems but are subleading at large $L$. 

By contrast, the two quantities $\langle\log_2|\vec{\Omega}_A|\rangle$ and $\log_2\langle |\vec{\Omega}_A|\rangle$ have markedly different behaviors below the critical temperature in the isotropic system. As discussed above, in this case the fluctuations around the saddle point change the scaling of $\langle\log_2|\vec{\Omega}_A|\rangle$ with $L$ but not the scaling of $\log_2\langle |\vec{\Omega}_A|\rangle$. 

In the inset of Fig.~\ref{fig:EDisotropic}(a) we show $\langle\log_2|\vec{\Omega}_A|\rangle$ versus $\log_2 L$ for various temperatures $T$ in the ordered phase. Our results are consistent with an increase of the form $\langle\log_2|\vec{\Omega}_A|\rangle = (8/3)\log_2 L + f_D(T)$ predicted in Eq.~\eqref{eq:isotropic_ordered}, where the nonuniversal function $f_D(T)$ is estimated via least-squares fit (see inset). Having extracted $f_D(T)$, in the main panel we find excellent agreement between our Monte Carlo data and Eq.~\eqref{eq:isotropic_ordered}. The decrease of $f_D(T)$ at small $T$ can be understood as a consequence of the phenomenological stiffness parameter $r$ [see Eq.~\eqref{eqn:standard_theory}] becoming large deep within the ordered phase. The behavior of $\log_2\langle |\vec{\Omega}_A|\rangle$ is significantly simpler [see Eq.~\eqref{eq:isotropic_ordered}]: even when the twist axis is in the plane of the entanglement cut, contributions from fluctuations around the twist are expected to give rise to subleading additive corrections. In Fig.~\ref{fig:EDisotropic}(b) and its inset, performing a similar analysis to that in Fig.~\ref{fig:EDisotropic}(a), we indeed find that $\log_2\langle |\vec{\Omega}_A|\rangle$ grows as $3 \log_2 L$ in the ordered phase. 

\begin{figure}
\includegraphics[width=\linewidth]{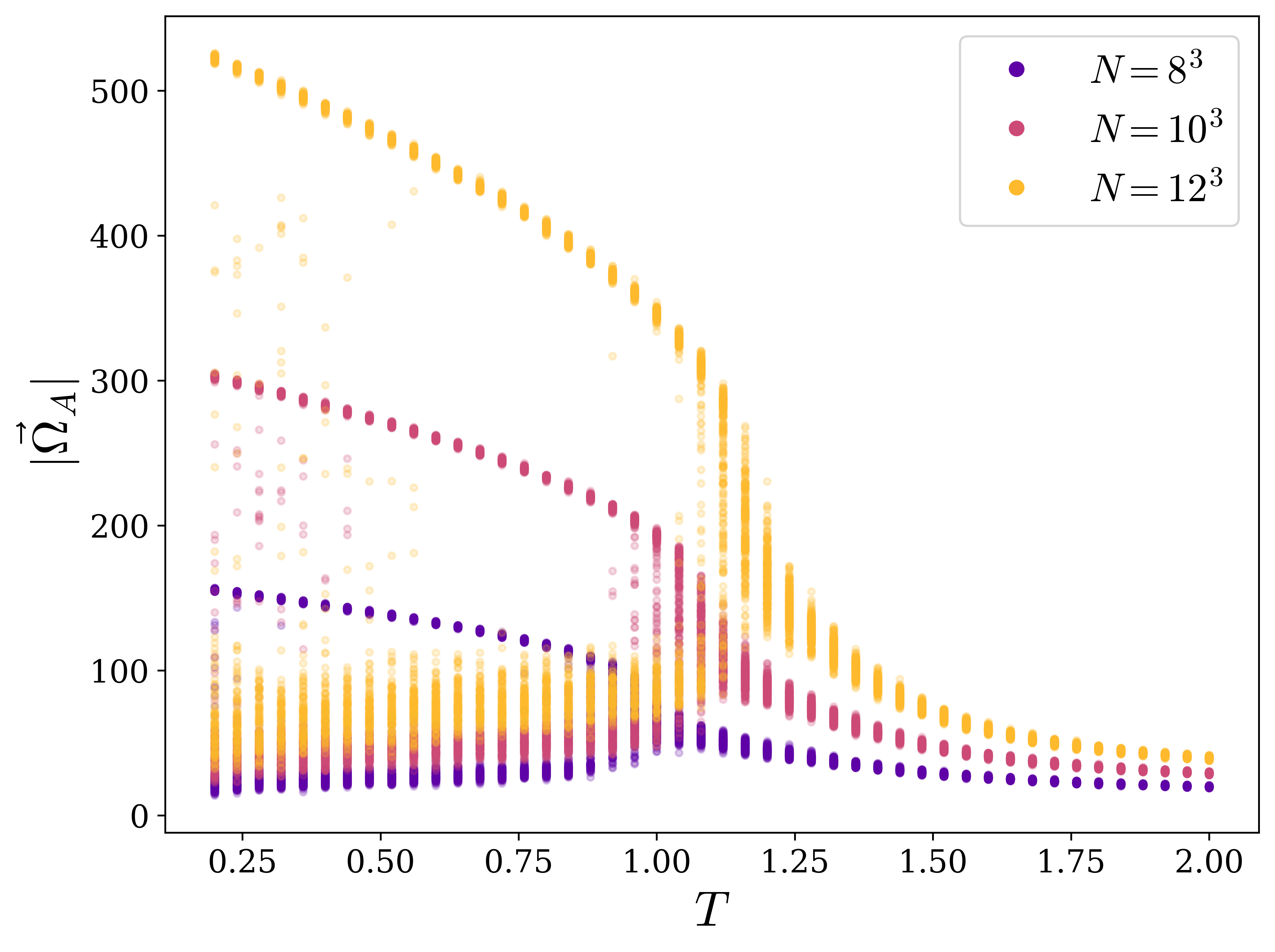}
\includegraphics[width=\linewidth]{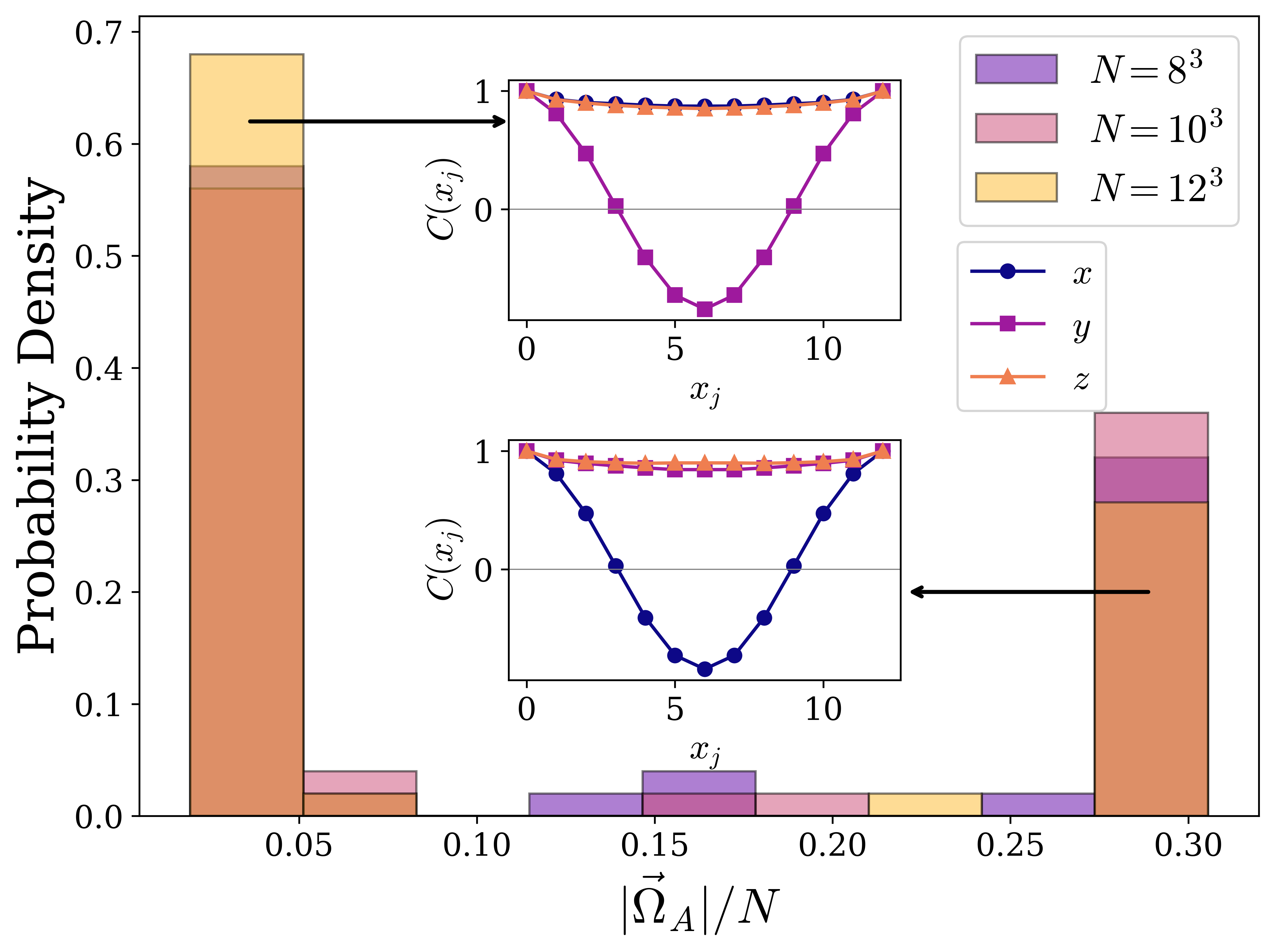}
\caption{Monte Carlo calculations of the distribution of $|\vec{\Omega}_A|$ for isotropic systems $L=L_{x,y,z}$ with $L = 8, 10, 12$ (legend) and $S = 50$. Here the bipartition between two half-systems $AB$ is fixed in the $y-z$ plane. For each system size, we run 50 independent annealing Monte Carlo calculations starting from high temperature. (a) shows $|\vec{\Omega}_A|$ of the 50 runs for each temperature, where every dot in the figure at a given temperature represents a $\langle |\vec{\Omega}_A|\rangle$ value from one run. (b) shows the histogram of $|\vec{\Omega}_A|/N$ at temperature $T = 0.2$, and the two insets are the spin-spin correlation function $C(x_j) = {\langle \hat{\Omega}(\vec{x}) \cdot \hat{\Omega}(\vec{x}+x_j \hat{e}_j)\rangle_{\mathcal{S}_0}}$, along three different lattice directions $\hat{e}_{j=x,y,z}$, computed for $L=12$ configurations randomly picked from the given bin as indicated.
}
\label{fig:omega_dist}
\end{figure}

To further probe the structure of low temperature saddle point in this geometry, in Fig.~\ref{fig:omega_dist}(a) we show the values of $|\vec{\Omega}_A|$ in 50 independent Monte Carlo calculations for each temperature, with the bipartition fixed in $y-z$ plane. We indeed see that the distribution of $|\vec{\Omega}_A|$ splits into two peaks when the temperature is lowered across the phase transition. 

A finer test comes from looking at the histogram of $|\vec{\Omega}_A|/N$ at the lowest temperature $T=0.2$, as shown in the main panel of Fig.~\ref{fig:omega_dist}(b). As expected, we find $\sim \frac{1}{3}$ of the weight in the histogram around $|\vec{\Omega}_A|\sim N$, with the remaining $\sim \frac{2}{3}$ of the weight concentrated around small $|\vec{\Omega}_A|$. Configurations drawn from the bin with $|\vec{\Omega}_A|\sim N$ have approximately cosinusoidal spin-spin correlations along the $\hat{x}$ direction and approximately constant correlations along $\hat y$ and $\hat z$. Configurations drawn from the smallest $|\vec{\Omega}_A|$ bin instead have approximately constant spin correlations along the $\hat{x}$ direction, and cosinusoidal correlations along one of the transverse directions (which in the figure we choose to identify with $\hat{y}$). This is in agreement with our picture of spin correlations in ordered isotropic systems. 

Through this section we have developed a theory for spin correlations in the $\rm{SU}(2)$ singlet sector of spin systems with geometrically local ferromagnetic interactions. Due to the connection between spin correlations and entanglement established in Sec.~\ref{sec:entanglement}, our results provide us with lower bounds on both the logarithmic entanglement negativity and the distillable entanglement. In Sec.~\ref{subsec:smiclass} we derived the explicit form of the singlet constraint, and then provided analytical arguments and numerical evidence (see Fig.~\ref{fig:f_integration}) that it can be approximated by a Gaussian suppression of the total spin. This allowed us to develop a continuum theory for spin correlations and arrive at concrete predictions for the behavior of various mixed-state entanglement measures in Sec.~\ref{subsec:IV.SCT_analysis}. We then confirmed these predictions using Monte Carlo simulations of the semiclassical theory in Sec.~\ref{subsec:monte_carlo}.

\section{Exact numerics in 1D}\label{sec:numerics}

The semiclassical theory developed through the previous section is controlled by large $S$, but we expect that our conclusions are more general. In particular, we expect that there is a finite-temperature entanglement transition in the singlet Gibbs state at arbitrary $S$. Unfortunately it is challenging to numerically probe the transition at small $S$ since it should only occur in spatial dimensions $d \geq 3$, and an exact representation of the singlet Gibbs state requires classical memory that is exponential in the number of spins. 

However, even working in $d=1$, we can test one of the central predictions of the theory, namely that the transition in mixed-state entanglement is tied to the divergence of the correlation length. In a disordered phase with a finite correlation length $\xi$, the lower bounds on distillable entanglement and the logarithmic negativity should each grow with $\xi$ and $L$ as $\frac{1}{2}\log_2 (L\xi)$ [Eq.~\eqref{eqn:disorder_scaling}]. In section \ref{subsec:1D_EE_numerics} we perform exact numerical calculations for the singlet Gibbs state of the spin$-1/2$ ferromagnetic Heisenberg chain and verify the predicted scaling. 

Following this, in section \ref{subsec:1D_gs_numerics} we show that the exact ground state of the 1D Heisenberg ferromagnet in the singlet sector closely resembles a twisted spin configuration, and so has correlations closely resembling the semiclassical spin configurations predicted in the ordered phase.

\subsection{Entanglement in 1D} \label{subsec:1D_EE_numerics}

We use exact diagonalization to calculate the lower bound on distillable entanglement $E^{D\;<}_{A:B}$ and the exact value of the logarithmic negativity $E^{N}_{A:B}$ in the singlet Gibbs state of spin$-1/2$ ferromagnetic Heisenberg chain.

Although the Heisenberg chain has no ordered phase at finite temperature, in a finite system the correlation length $\xi$ can grow until it becomes comparable to the system size $L$, producing effective long-range order. The semiclassical theory in Sec.~\ref{subsec:IV.SCT_analysis} predicts that the lower bounds for both distillable entanglement and logarithmic negativity scale as $\frac{1}{2}\log(L\xi)$ in the paramagnetic phase, so this crossover in $\xi$ translates directly into a crossover in entanglement scaling. When $\xi \ll L$, the system behaves as a paramagnet and the entanglement bounds reduce to $\log\sqrt{L}$. When the temperature is lowered to reach the regime $\xi \sim L$, effective ferromagnetic ordering sets in and the bounds scale as $\log L$. Since the correlation length scales linearly with inverse temperature $\xi \sim \beta$ in the $d=1$ ferromagnetic Heisenberg model~\cite{Arovas19881DCorr}, the crossover between these regimes occurs at $\beta \sim L$. We therefore expect a universal crossover in $E^{D\;<}_{A:B}$ and $E^{N\;<}_{A:B}$ from $\log L $ to $\log\sqrt{L}$ as $\beta \sim L$.

\begin{figure}
\includegraphics[width=0.95\linewidth]{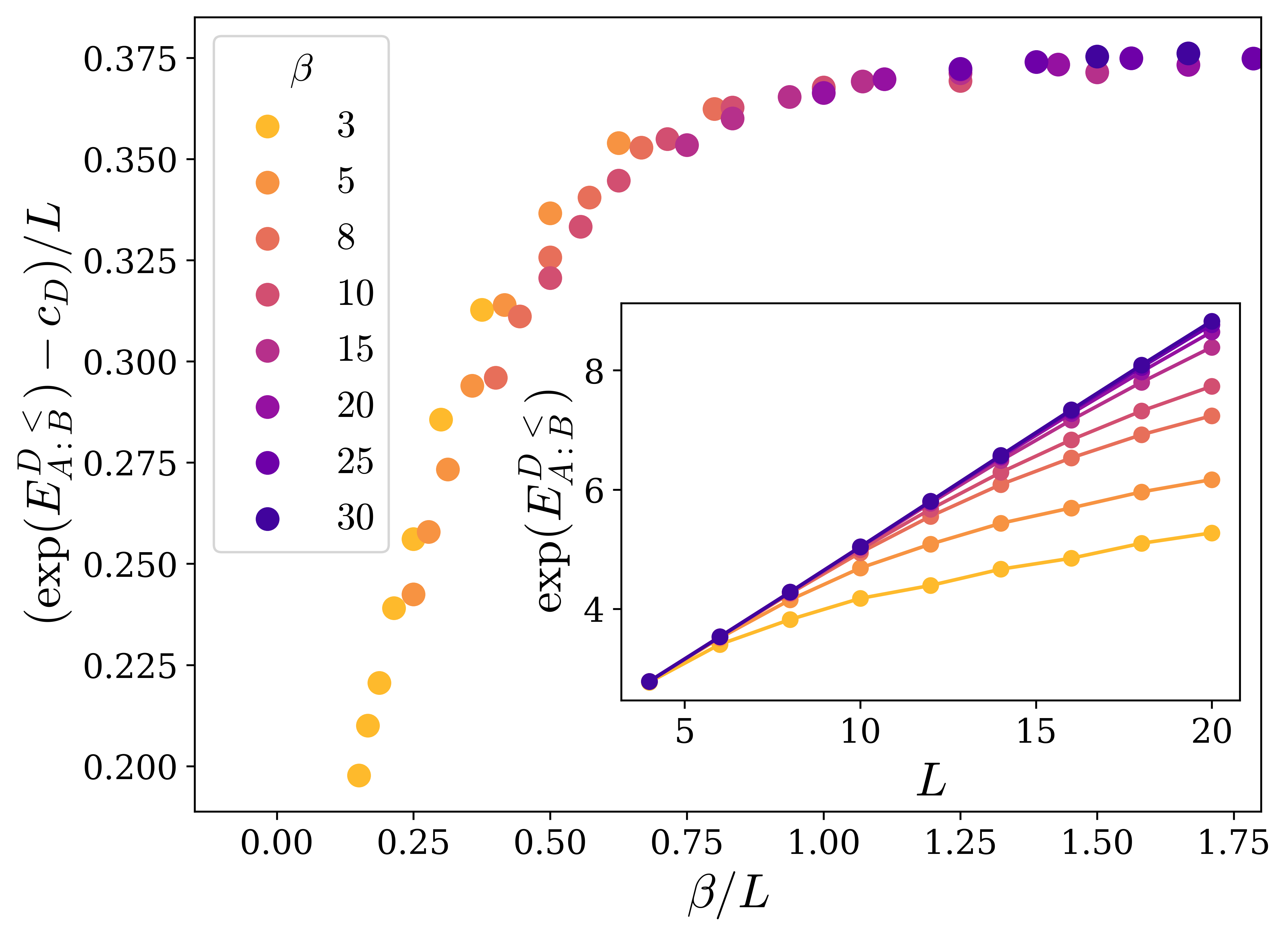}
\includegraphics[width=0.95\linewidth]{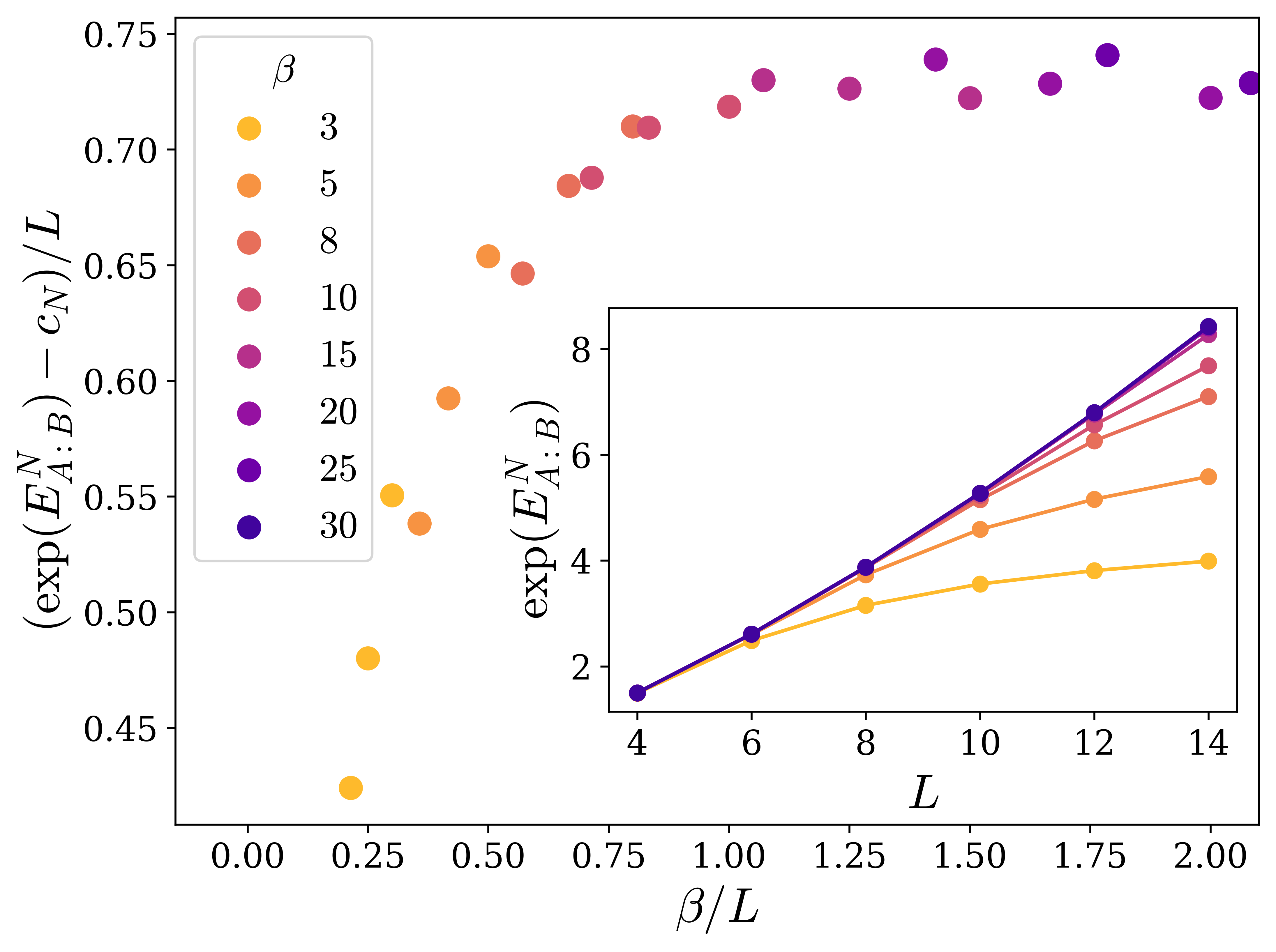}
\caption{Exact diagonalization calculations of bipartite (a) distillable entanglement lower bound and (b) logarithmic negativity of 1D spin-1/2 ferromagnetic Heisenberg model's Gibbs state projected onto total spin singlet sector, for various temperatures (see legend). In the inset of (a) we show the scaling of the exponential of distillable entanglement lower bound $\exp(E^{D\;<}_{A:B})$ for different temperatures, plotted against system size $L$. The main panel of (a) shows the scaling collapse for $\exp(E^{D\;<}_{A:B})$ of the form Eq.\eqref{eq:scaling_collapse}, where the constant $c_D$ is determined by the vertical intercept of the linear fitting to $\exp(E^{D\;<}_{A:B})$ against $L$, using the lowest temperature curve. We expect all data points to collapse onto the same curve which increases algebraically with $\beta/L$ when $\beta/L\ll1$, and plateau when $\beta/L\gg1$. In the inset of (b) we show the scaling of the exponential of logarithmic negativity $\exp(E^{N}_{A:B})$ for different temperatures, plotted against system sizes $L$. The main panel of (b) shows the scaling collapse of $\exp(E^N_{A:B})$ using the scaling form expect to collapse the lower bound $\exp(E^{N\;<}_{A:B})$. The constant $c_N$ is determined by the vertical intercept of the linear fitting to $\exp(E^{N}_{A:B})$ against $L$, using the lowest temperature curve.}
\label{fig:exact_entanglement}
\end{figure}

The numerical results for $E^{D\;<}_{A:B}$ and $E^N_{A:B}$ are shown in Figs.~\ref{fig:exact_entanglement}(a) and \ref{fig:exact_entanglement}(b), respectively. For ease of visualization and comparison, we plot the exponential of these quantities, and they are shown in the insets of these figures. We see that both $\exp(E^{D\;<}_{A:B})$ and $\exp(E^N_{A:B})$ scale linearly with system size $L$ for sufficiently large $\beta$, and sublinearly for small $\beta$.

To understand this, note that in $d=1$ our previous analysis suggests $\exp(E^{D\;<}_{A:B}) \approx A_D\sqrt{\xi L}+c_D$, where $A_D$ and $c_D$ are $\mathcal{O}(1)$ constants. Equivalently,
\begin{equation}\label{eq:scaling_collapse}
    \left(\exp(E^{D\;<}_{A:B}) - c_D\right)/L \approx A_D\sqrt{\xi/L}.
\end{equation}
First, when $\beta/L \gg 1$, the correlation length $\xi$ saturates at $L$ so the left-hand side of Eq.~\eqref{eq:scaling_collapse} is a constant with respect to $\beta/L$. When $\beta/L \ll 1$, using $\xi \propto \beta$, Eq.~\eqref{eq:scaling_collapse} instead scales as $\sqrt{\beta/L}$. Therefore we expect  $\left(\exp(E^{D\;<}_{A:B}) - c_D\right)/L$ plotted against $\beta/L$ to collapse for all $\beta$ and $L$. Specifically, we expect this quantity to increase algebraically with $\beta/L\ll 1$, and plateau when $\beta/L\gg1$. 

This collapse is shown in the main panel of Fig.~{\ref{fig:exact_entanglement}(a)}. There we have extracted the constant $c_D$ from the vertical intercept of a linear fit to $\exp(E^{D\;<}_{A:B})$ versus $L$, using data at the smallest temperature shown. The collapse confirms that the growth of entanglement in the singlet sector is tied to the divergence of the spin correlation length. 

We also test the same scaling form [Eq.~\eqref{eq:scaling_collapse}] for $E^N_{A:B}$, with different constants $A_N$ and $c_N$. Similarly, the constant $c_N$ is determined by the vertical intercept of a linear fit to $\exp(E^{N}_{A:B})$ against $L$, using the lowest temperature data. The result is shown in the main panel of Fig~\ref{fig:exact_entanglement} (b). Despite the fact that our analysis directly predicts the scaling of $E^{D\;<}_{A:B}$ and $E^{N\;<}_{A:B}$ which only bounds $E^N_{A:B}$ from below, the collapse for $E^N_{A:B}$ agrees well with our prediction.

\subsection{Ground State in 1D}\label{subsec:1D_gs_numerics}

Here we study the ground states of ferromagnetic Heisenberg chains with $S=1/2$ and $S=1$, and compare them with an ansatz inspired by our semiclassical theory. 

Consider a simple twisted spin configuration on a chain:
\begin{equation}\label{eqn:1D_saddle}
\begin{split}
    \bar{\Omega}_j = \left(\sin\frac{2 \pi j}{L},0,\cos\frac{2 \pi j}{L}\right).
\end{split}
\end{equation}
Projecting the corresponding spin coherent state $\bigotimes_j \ket{\bar{\Omega}_j}$ into the total singlet sector produces an ansatz for the ground state in the singlet sector:
\begin{equation} \label{eqn:gs_ansatz}
    \ket{\text{GS}} \propto \Pi_0 \bigotimes_j \exp\left(i \frac{2 \pi j}{L} {S}_j^y\right) \ket{z}_j,
\end{equation}
where $S^z_j\ket{z}_j = S\ket{z}_j$. Note that although in Eq. \eqref{eqn:1D_saddle} the twist is in the $S^x-S^z$ plane, choosing the twist to be in other planes doesn't affect the quantum state Eq.~\eqref{eqn:gs_ansatz}. This is because the different possible twists are related to each other by global rotations, so they will all be projected onto the same state by $\Pi_0$.

\begin{figure}
\includegraphics[width=0.95\linewidth]{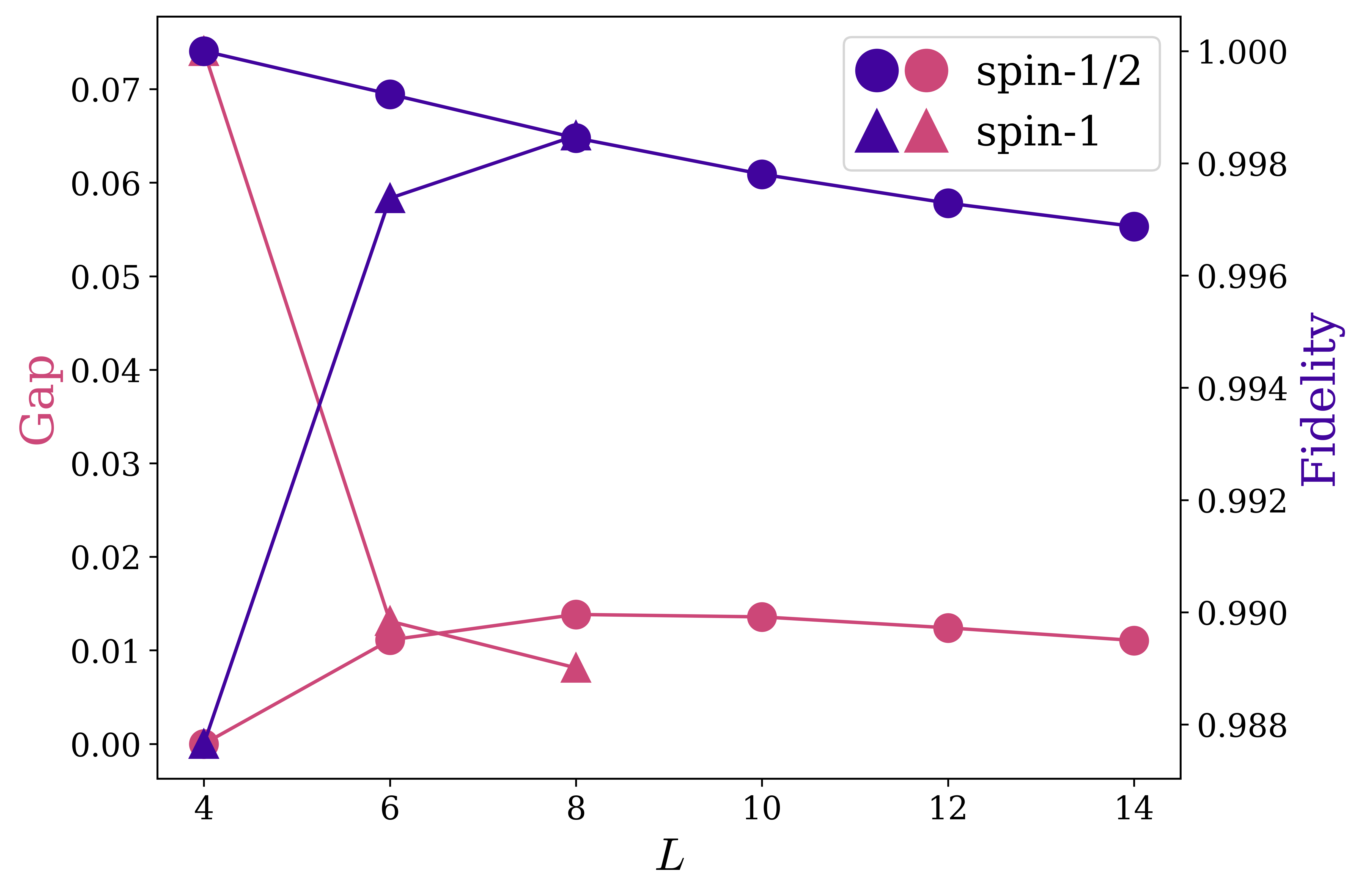}
\caption{Energy gap (pink curves, scale on the left) and fidelity (blue curves, scale on the right) between 1D singlet ferromagnetic Heisenberg model ground state ansatz Eq. \eqref{eqn:gs_ansatz} and true ground state obtained from exact diagonalization, plotted for various system sizes $L$. The circles represent spin-1/2 and triangles represent spin-1.
}
\label{fig:exact_ground_state}
\end{figure}

In Fig.~\ref{fig:exact_ground_state} we numerically evaluate (i) the energy gaps between our ansatz and the true ground states of Heisenberg chains in the singlet sector and (ii) the fidelities between these states. We use periodic boundary conditions, various $L$, and $S=\frac{1}{2},1$. Remarkably we find that the fidelities for both $S$ are above $0.98$ for all $L$, and, moreover, that they are still above $0.99$ even for $L=14$ in the $S=1/2$ case. The energy gaps are correspondingly small, being well below $L^{-1}$ for the system sizes that we study. 

At this stage we do not have a rigorous analytical relation between our ansatz and the true ground state; we leave the study of this to future works.

\section{Discussion} \label{discussion}

A central challenge in many-body physics is to identify intrinsically quantum-mechanical phenomena that survive in equilibrium. In this work we have shown that, in systems with strong $\rm{SU}(2)$ symmetry, finite-temperature ordering transitions are typically associated with entanglement transitions. These phenomena arise in singlet Gibbs states, non-Abelian generalizations of canonical ensembles which maximize entropy subject to the constraints of a fixed energy expectation value and global $\rm{SU}(2)$ invariance. 

First, building on Refs.~\cite{Li2025HighlyEntangled,Moharramipour2024MMIS}, we derived relations between measures of mixed-state entanglement and spin correlations that apply quite generally to systems with strong $\rm{SU}(2)$ symmetry. These led us to develop a theory for correlations in singlet Gibbs states based on spin coherent state path integrals. In the semiclassical limit of large $S$, where correlations can be expressed as statistical averages over spatial configurations of semiclassical spins (three-component unit vectors), these theories encode the singlet constraint through a complicated function of the full configuration. 

We then argued that, at high temperatures and in the vicinity of continuous thermal phase transitions, this constraint simplifies dramatically and takes the form of a Gaussian suppression of the total semiclassical spin. By evaluating the exact semiclassical action (which involves the singlet constraint) we also provided extensive numerical evidence for the validity of this Gaussian approximation. 

Coarse-graining the resulting theory we arrived at an analytically tractable description of the finite-temperature ordering transition in the singlet sector, and therefore of the entanglement transition. In the disordered phase we showed that the lower bounds on the half-system distillable entanglement and the logarithmic entanglement negativity scale as $(d/2)\log(L \xi)$, where $\xi$ is the correlation length. In a ferromagnetic ordered phase these mixed-state entanglement measures have additional contributions from Goldstone modes and the local background magnetization. As a consequence we find that these quantities each scale as $\log L$ with $d$-dependent prefactors that are larger than in the paramagnetic phase; on decreasing the temperature these prefactors undergo $\mathcal{O}(1)$ increases over a temperature interval that is narrow at large $L$. We have supported this analytical result with Monte Carlo simulations of our semiclassical lattice model in three spatial dimension. There we found clear signatures of the transition in properties of the spin configurations which, in the singlet sector, are proxies for mixed-state entanglement measures. 

To provide further support for our theory away from the semiclassical limit of large $S$, we numerically studied mixed-state entanglement in the singlet Gibbs state of a one-dimensional spin $S=1/2$ Heisenberg ferromagnet. There is no finite-temperature entanglement transition in $d=1$, but the correlation length $\xi$ diverges to the scale of the system length $L$ at an inverse temperature $\beta \sim L$. Through exact numerical calculations of our lower bound on distillable entanglement, and of the exact logarithmic negativity, we numerically confirmed a prediction of our semiclassical theory that these quantities should scale as $(d/2)\log(L \xi)$ when $L \xi$ is large. 

Immediate next questions concern the fate of mixed-state entanglement in the singlet Gibbs states of antiferromagnetic insulators, and also in systems with itinerant degrees of freedom, such as Stoner ferromagnets \cite{Auerbach1994Spins}. More exotic still are Gibbs states that are singlets under different non-Abelian symmetries. For example it is interesting to ask to what extent the behavior we have identified carries over to total singlets under ${\rm{SU}}(n)$ with $n > 2$. Such an investigation could also provide new insights into the behavior of $\rm{SU}(2)$ singlets at small $S$, complementing the large-$S$ theory developed above.

Our work has provided an example of a family of entanglement transitions in the steady states of local Lindbladians. This is because singlet Gibbs states arise from $\rm{SU}(2)$-symmetric interactions between the system and a thermal environment, and such interactions can be recast as system Lindbladians whose individual terms act on subregions of linear extent $\sim \beta$~\cite{Chen2023Gibbs}.

It would be interesting to determine the time required to generate singlet Gibbs states. Even at infinite temperature these states have distillable entanglement $\approx (d/2)\log L$, and under local open-system dynamics it must take a time at least $\sim L$ to generate this entanglement. By contrast, the infinite-temperature state without any symmetries can of course be created in time of order unity. In the case of a Gibbs state with strong $\rm{U}(1)$ symmetry, the bottleneck to thermal state preparation is the diffusion of the $\rm{U}(1)$ symmetry charge. However, it is unclear how the time scale for the emergence of mixed-state entanglement in systems with strong non-Abelian symmetry is related to charge transport. One can ask similar questions in open quantum systems exhibiting Hilbert-space fragmentation, whose steady states can even be volume-law entangled \cite{Li2025HighlyEntangled}.

An intriguing byproduct of our analysis is an ansatz for the ground state of the singlet ferromagnetic Heisenberg chain. We have found a surprisingly large fidelity with the true ground state for $S=\frac{1}{2}$ and $S=1$, but the reason for this requires further investigation. Equally interesting is the ground state structure of the singlet ferromagnetic Heisenberg model in higher dimensions. One could ask, for example, whether projecting a simple twisted configuration onto the singlet sector remains a good ansatz beyond one spatial dimension.

\acknowledgements

The authors are grateful to Zack Weinstein and Pablo Sala for useful discussions. This work was supported by the Gordon \& Betty Moore Foundation (SJG), the NSF QLCI
program through Grant No. OMA-2016245 (EA), and a
Simons Investigator Award (EA). Numerical calculations were done using services provided by the OSG Consortium \cite{OSPool, OpenScienceDataFed, osg07, osg09}, which is supported by the National Science Foundation awards 2030508 and 2323298. 


\bibliography{apssamp}

\appendix
\onecolumngrid

\section{Derivation of the Singlet Constraint in the Semiclassical Theory} \label{app:int_over_U}

In this Appendix we show how the singlet constraint enters into a semiclassical (large $S$) path integral, or into a $\mathrm{SU}(2)$-symmetrized separable state (which, after symmetrization, is no longer separable). Consider a many-spin state of the form
\begin{align}
	\rho = \int d\hat{\bm{\Omega}}\, w(\hat{\bm{\Omega}}) \ket{\hat{\bm{\Omega}}}\bra{\hat{\bm{\Omega}}}.
\end{align}
Up to normalization, we can project such a state into the singlet sector via
\begin{align}
	\int dU_1 dU_2 \, U_1^{\otimes N} \rho U_2^{\otimes N}.
\end{align}
Here $dU_1$ and $dU_2$ are Haar measures for the spin-$S$ representation of $\mathrm{SU}(2)$. The corresponding partition function is $\mathcal{Z}=\int dU \text{Tr}[U^{\otimes N}\rho]$, where we have used Haar invariance to combine the integrals over the global $\mathrm{SU}(2)$ rotations. Rewriting the trace as a coherent-state path integral then gives
\begin{equation}
    \mathcal{Z}=\int d\hat{\bm{\Omega}}\,w(\hat{\bm{\Omega}})\int dU \bra{\hat{\bm{\Omega}}} U^{\otimes N} \ket{\hat{\bm{\Omega}}}.
\end{equation}
So, $C(\bm{\hat{\Omega}}) \equiv - \ln \int dU  \bra{\bm{\hat{\Omega}}} U^{\otimes N}\ket{\bm{\hat{\Omega}}}$ can be viewed as a contribution to the action that implements the singlet constraint. Below we calculate this contribution. 

\subsection{Simplification of the Integrand}

We first simplify the expectation value $\bra{\bm{\hat{\Omega}}} U^{\otimes N}\ket{\bm{\hat{\Omega}}}$ by writing $\ket{\hat{\bm{\Omega}}} = \bigotimes_j\ket{\hat{\Omega}_j}= \bigotimes_jV_j\ket{S,S}_j$, where $\ket{S,S}_j$ denotes a standard angular momentum eigenstate at site $j$, i.e. with $S^z_j \ket{S,S}_j = S\ket{S,S}_j$, and $V_j$ is a unitary operator at $j$. Additionally note that any $\mathrm{SU}(2)$ rotation can be parametrized in the angle-axis representation $U^{\otimes N} = e^{i \gamma \hat{n} \cdot \vec{S}_{\text{tot}}}$. The integrand can then be expressed as
\begin{equation}\label{eqn:sc_onesite}
\begin{split}
    & \bra{\hat{\bm{\Omega}}}U^{\otimes N}\ket{\hat{\bm{\Omega}}}  = \prod_j \bra{\hat{\Omega}_j}U_j\ket{\hat{\Omega}_j}  = \prod_j \bra{S,S}V_j^\dagger e^{i \gamma \hat{n} \cdot \vec{S}_{j}}V_j\ket{S,S} = \prod_j \bra{S,S}e^{i \gamma (R_j\hat{n}) \cdot \vec{S}_{j}}\ket{S,S},\\
\end{split}
\end{equation}
where $R_j$ is a $3 \times 3$ orthogonal rotation matrix fixed by $V_j$, and $R_j\hat{n}$ is the axis of the $\mathrm{SU}(2)$ rotation acting on $\ket{S,S}$. 

Now each $e^{i \gamma (R_j\hat{n}) \cdot \vec{S}_{j}}\ket{S,S}_j$ is a new one-site spin coherent state, and any such state can be expanded in the angular momentum basis as \cite{Auerbach1994Spins}:
\begin{equation}\label{eqn:rot_sc_angular}
\begin{split}
    e^{i \gamma (R_j\hat{n}) \cdot \vec{S}_{j}}\ket{S,S}_j & = e^{iS\chi_j}\sqrt{(2S)!}\sum_m \frac{u_j^{S+m}v_j^{S-m}}{\sqrt{(S+m)!(S-m)!}}\ket{S,m}_j,
\end{split}
\end{equation}
where $u_j$, $v_j$, and $\chi_j$ are $\gamma$- and  $\left(R_j\hat{n}\right)$- dependent factors determined by the transformation rules of $\mathrm{SU}(2)$ Schwinger bosons. Specifically \cite{Auerbach1994Spins}:
\begin{equation}
    e^{i \gamma (R_j\hat{n}) \cdot \vec{S}_{j}}\left(
    \begin{smallmatrix}
    a^\dagger\\b^\dagger
    \end{smallmatrix}\right) e^{-i \gamma (R_j\hat{n}) \cdot \vec{S}_{j}} = e^{i \frac{\gamma}{2}(R_j\hat{n})\cdot \vec{\sigma}}\left(
    \begin{smallmatrix}
    a^\dagger\\b^\dagger
    \end{smallmatrix}\right) \equiv \left(
    \begin{smallmatrix}
    u_j e^{i\chi_j/2} & v_j e^{i\chi_j/2}\\-v_j^* e^{-i\chi_j/2} & u_j^* e^{-i\chi_j/2}
    \end{smallmatrix}\right)\left(
    \begin{smallmatrix}
    a^\dagger\\b^\dagger
    \end{smallmatrix}\right),
\end{equation}
with $\vec{\sigma} = (\sigma^x, \sigma^y, \sigma^z)$ being Pauli matrices.
Requiring the matrix exponential $e^{i \frac{\gamma}{2}(R_j\hat{n})\cdot \vec{\sigma}}$ to match entry-wise with the $2 \times 2$ matrix above gives the following relation:
\begin{equation}\label{eqn:integrand_relation}
    u_je^{i\chi_j/2} = \cos\frac{\gamma}{2}+i(\hat{n}\cdot \hat{\Omega}_j)\sin\frac{\gamma}{2}
\end{equation}

Therefore, substituting Eq.~\eqref{eqn:rot_sc_angular} into Eq.~\eqref{eqn:sc_onesite} gives
\begin{equation}\label{ExactWeight}
\begin{split}
    \bra{\hat{\bm{\Omega}}}U^{\otimes N}\ket{\hat{\bm{\Omega}}}  &= \prod_j \bra{S,S}e^{iS\chi_j}\sqrt{(2S)!}\sum_m \frac{u_j^{S+m}v_j^{S-m}}{\sqrt{(S+m)!(S-m)!}}\ket{S,m}= \prod_j e^{iS\chi_j}u_j^{2S},
\end{split}
\end{equation}
and Eq.~\eqref{eqn:integrand_relation} further simplifies it to
\begin{equation}\label{eqn:simplified_integrand}
\begin{split}
    \bra{\hat{\bm{\Omega}}}U^{\otimes N}\ket{\hat{\bm{\Omega}}}     &  = \prod_j \left(\cos\frac{\gamma}{2}+i\sin\frac{\gamma}{2}\hat{n}\cdot \hat{\Omega}_j\right)^{2 S} = \exp\left(2SN \Bigg[\frac{1}{N}\sum_j \log(\cos\frac{\gamma}{2}+i\sin\frac{\gamma}{2}\hat{n}\cdot \hat{\Omega}_j)\Bigg]\right)
\end{split}
\end{equation}

\subsection{Saddle Point Approximation for $C(\hat{\bm{\Omega}})$}
The form of the exponent in Eq.~\eqref{eqn:simplified_integrand} hints that in the large $SN$ limit, the integral can be evaluated using a saddle point approximation. We will focus on non-collinear configurations, since collinear configurations are entropically disfavored. For non-collinear configurations the integrand has a unique saddle point at $\gamma = 0$. So we expand the exponent to quadratic order in $\gamma$, and allow for width $1/\sqrt{NS}$ fluctuations. With this, and the $\mathrm{SU}(2)$ Haar measure in the angle-axis parametrization $\int dU = \int d\hat{n} \frac{1}{\pi}\sin^2\frac{\gamma}{2}d\gamma$ \cite{DAndrea2024SU2Haar}, the integral over $U$ becomes:
\begin{equation}\label{eqn:Wderivation}
\begin{split}
    & e^{-C(\hat{\bm{\Omega}})} = \int dU \bra{\hat{\bm{\Omega}}} U^{\otimes N} \ket{\hat{\bm{\Omega}}}  = \int d\hat{n}\frac{1}{\pi}\sin^2\frac{\gamma}{2}d\gamma \exp\left(2NS \frac{1}{N} \sum_j \log(\cos\frac{\gamma}{2}+i\sin\frac{\gamma}{2}\hat{n}\cdot \hat{\Omega}_j)\right)\\
    &\approx \int d\hat{n} \int d\gamma \frac{\gamma^2}{4\pi}\exp\left(2NS \frac{1}{N} \sum_j\left(\frac{1}{8}\left(-1+(\hat{n}\cdot \hat{\Omega}_j)^2\right)\gamma^2+i\frac{\gamma}{2}\hat{n} \cdot \hat{\Omega}_j\right)\right)\\
    & = \int d\hat{n}\;\; \exp\left(\frac{S}{\sum_j (\hat{n}\cdot \hat{\Omega}_j)^2-N}(\hat{n} \cdot \vec{\Omega}_{\text{tot}})^2\right)\frac{\pi^{-1/2} \left(2 S (\hat{n} \cdot \vec{\Omega}_{\text{tot}})^2+\sum_j (\hat{n}\cdot \hat{\Omega}_j)^2-N\right)\sqrt{S\left(-\sum_j (\hat{n}\cdot \hat{\Omega}_j)^2+N\right)}}{S^2\left(\sum_j (\hat{n}\cdot \hat{\Omega}_j)^2-N\right)^3},
\end{split}
\end{equation}
where in the last step we perform a Gaussian integral over $\gamma$.

The residual integral over the unit vector $\hat{n}$ is difficult to evaluate exactly. However, as explained in Sec.\ref{subsec:smiclass} of the main text, as long as the ordered moment is small (or zero) the spatial averages appearing in the denominators can be simplified. Specifically, $a[\bm{\hat{\Omega}}] \equiv 1-\frac{1}{N}\sum_j (\hat{n} \cdot \hat{\Omega}_j)^2 \approx \frac{2}{3}$ up to corrections that are subleading when the system size is large and the total magnetization is small. 

Substituting constant $a \approx \frac{2}{3}$ into Eq.~\eqref{eqn:Wderivation} gives:
\begin{equation}
\begin{split}
    e^{-C(\hat{\bm{\Omega}})}  & \approx \int d\hat{n}\;\; \exp\left(-\frac{S}{aN}(\hat{n} \cdot \vec{\Omega}_{\text{tot}})^2\right)\frac{\pi^{-1/2} \left(aN - 2 S (\hat{n} \cdot \vec{\Omega}_{\text{tot}})^2\right)\sqrt{aSN}}{a^3S^2N^3}\\
    & = \frac{\pi^{-1/2} \sqrt{aSN}}{a^3S^2N^3}\int d\left(\cos\theta\right)\frac{d\phi}{4\pi}\; \exp\left(-\frac{S}{aN}(|\vec{\Omega}_{\text{tot}}|\cos\theta)^2\right)\left(aN - 2 S (|\vec{\Omega}_{\text{tot}}|\cos\theta)^2\right)\\
    & = \frac{\pi^{-1/2}}{(aSN)^{3/2}} e^{-\frac{S}{aN}\vec{\Omega}_{\text{tot}}^2}.
\end{split}
\end{equation}
Note that the compact integral over $\cos\theta$ can be evaluated exactly by identifying the integrand as derivative of $aN\cos\theta e^{-\frac{S}{aN}(|\vec{\Omega}_{\text{tot}}|\cos\theta)^2}$ with respect to $\cos\theta$.

Keeping only what's relevant for understanding the effective free energy coming from the singlet projector, and substituting in $a \approx \frac{2}{3}$ we arrive at the constraint for typical spin configurations:
\begin{equation}\label{Wfinal}
    C(\hat{\bm{\Omega}}) \approx \frac{3}{2}\frac{S}{N}\vec{\Omega}_{\text{tot}}^2 + \ldots,
\end{equation}
where the ellipsis represents constant contributions that do not alter correlations.

\section{The Correlation Function In the Semiclassical Theory} \label{app:correlation_function}

In this Appendix, we provide technical details in obtaining the spin-spin correlation function of the singlet Gibbs state represented using the semiclassical theory given by Eq. \eqref{eqn:semiclassical_corr}.

Starting from the path integral representation of the singlet Gibbs state density matrix 
\begin{align}
    \rho_0 \propto \int dU U^{\otimes N} e^{-\beta H} \propto \int dU U^{\otimes N} \ket{\hat{\bm{\Omega}}(\beta)}e^{-\mathcal{A}[\hat{\bm{\Omega}}(\tau)]}\bra{\hat{\bm{\Omega}}(0)},
\end{align}
where $\mathcal{A}[\hat{\bm{\Omega}}(\tau)]$ is given by Eq.~\eqref{eqn:SHeis}, we consider the same approximation used in the main text for the partition function that, in large $S$ limit, $\tau-$independent histories $\hat{\bm{\Omega}}(\tau) = \hat{\bm{\Omega}}(0)$ dominate the path integral. Using this, the spin-spin correlation function in the singlet sector can be expressed as
\begin{equation} \label{eqn:semiclassical_corr_app}
\begin{split}
    \mathrm{Tr}\left[\rho_0\vec{S}_i \cdot \vec{S}_j\right] \approx \frac{1}{\mathcal{Z}_0}\int d\hat{\bm{\Omega}}\int dU\bra{\hat{\bm{\Omega}}}\vec{S}_i \cdot \vec{S}_jU^{\otimes N}\ket{\hat{\bm{\Omega}}}e^{-\beta E[\hat{\bm{\Omega}}]},
\end{split}
\end{equation}
where $\mathcal{Z}_0$ is given by Eq.~\eqref{eqn:semiclassical_Z} and $E[\hat{\bm{\Omega}}]$ is the energy part of the action.

\subsection{Simplification of the Correlation Function}

Below we simplify the factor $\bra{\hat{\bm{\Omega}}}\vec{S}_i \cdot \vec{S}_jU^{\otimes N}\ket{\hat{\bm{\Omega}}}$ in the correlation function. For simplicity we abbreviate it as $\bra{\hat{\bm{\Omega}}}\vec{S}_i \cdot \vec{S}_j\ket{\hat{\bm{\Omega}}'}$, where $\ket{\hat{\bm{\Omega}}'} = U^{\otimes N}\ket{\hat{\bm{\Omega}}}$. Since $\ket{\hat{\bm{\Omega}}}$ and $\ket{\hat{\bm{\Omega}}'}$ are both product states, we only need to calculate ${\bra{\hat{\Omega}}\vec{S}\ket{\hat{\Omega}'}}$ where $\ket{\hat{\Omega}}$ and $\ket{\hat{\Omega}'}$ each are one-site spin coherent state labeled by their angles ${\theta, \phi, \theta', \phi'}$.

It is helpful to work in the angular momentum basis $\ket{S,p}$, with $\vec{S}^2\ket{S,p} = S(S+1)\ket{S,p}$ and  $S^z\ket{S,p} = (S-p)\ket{S,p}$, such that the basis states are orthonormal. The spin coherent state can be expanded in the angular momentum basis as \cite{Radcliffe1971CoherentState}:
\begin{equation}\label{eqn:sc_in_angularmomentum}
    \ket{\hat{\Omega}} = (1+|\mu|^2)^{-S}\sum_{p=0}^{2S}\left(\frac{(2S)!}{p!(2S-p)!}\right)^{1/2}\mu^p\ket{S,p},
\end{equation}
where $\mu = \tan\frac{\theta}{2}e^{i\phi}$. With $\vec{S} = (S^x, S^y, S^z)$, we can use Eq.~\eqref{eqn:sc_in_angularmomentum} to obtain each vector component of ${\bra{\hat{\Omega}}\vec{S}\ket{\hat{\Omega}'}}$.

The $z$ component is given by
\begin{equation}
\begin{split}
    & \bra{\hat{\Omega}}{S}^z\ket{\hat{\Omega}'} = (1+|\mu|^2)^{-S}(1+|\mu'|^2)^{-S}\sum_{p=0}^{2S} \frac{(2S)!}{p!(2S-p)!}\mu^{*p}\mu'^{p}(S-p)\\
    & = (1+|\mu|^2)^{-S}(1+|\mu'|^2)^{-S}\left(S\sum_{p=0}^{2S} \frac{(2S)!}{p!(2S-p)!}(\mu^{*}\mu')^{p}-\sum_{p=0}^{2S} \frac{(2S)!}{p!(2S-p)!}(\mu^{*}\mu')\frac{\partial}{\partial(\mu^{*}\mu')}(\mu^{*}\mu')^{p}\right)\\
    & = S\bra{\hat{\Omega}}\hat{\Omega'}\rangle \frac{\cos\theta+\cos\theta'-i\sin\theta\sin\theta'(\sin\phi'\cos\phi-\sin\phi\cos\phi')}{1+\cos\theta\cos\theta'+\sin\theta\sin\theta'\cos(\phi-\phi')}\\
    & = S\bra{\hat{\Omega}}\hat{\Omega'}\rangle\frac{{\Omega}^z+{\Omega}'^z-i{\Omega}^x{\Omega}'^y+i{\Omega}^y{\Omega}'^x}{1+\hat{\Omega} \cdot \hat{\Omega}'},\\
\end{split}
\end{equation}
where from the second line to the third line we've used binomial formula to simplify the sum over $p$, and that $\bra{\hat{\Omega}}\hat{\Omega'}\rangle = (1+|\mu|^2)^{-S}(1+|\mu'|^2)^{-S}(1+\mu^*\mu')^{2S}$.

To obtain the $x$ and $y$ components, it's useful to write $S^x = \frac{S^++S^-}{2}$ and $S^y = \frac{S^+-S^-}{2i}$, and we can use anologous method to obtain the matrix elements for $S^+, S^-$: 
\begin{align}
&\bra{\hat{\Omega}}{S}^+\ket{\hat{\Omega'}} = 2S\bra{\hat{\Omega}}\hat{\Omega'}\rangle\frac{\cos^2\frac{\theta}{2}\sin\theta'e^{i\phi'}+\sin^2\frac{\theta'}{2}\sin\theta e^{i\phi}}{1+\cos\theta\cos\theta'+\sin\theta\sin\theta'\cos(\phi-\phi')},\\
    & \bra{\hat{\Omega}}{S}^-\ket{\hat{\Omega'}} = \left(\bra{\hat{\Omega'}}{S}^+\ket{\hat{\Omega}}\right)^\dagger = 2S\bra{\hat{\Omega}}\hat{\Omega'}\rangle\frac{\cos^2\frac{\theta'}{2}\sin\theta e^{-i\phi}+\sin^2\frac{\theta}{2}\sin\theta' e^{-i\phi'}}{1+\cos\theta\cos\theta'+\sin\theta\sin\theta'\cos(\phi-\phi')}.
\end{align}
This gives the $x,y$ components:
\begin{align}
    \bra{\hat{\Omega}}{S}^x\ket{\hat{\Omega'}} & = \bra{\hat{\Omega}}\frac{{S}^+ +{S}^-}{2}\ket{\hat{\Omega}'} = S\bra{\hat{\Omega}}\hat{\Omega}'\rangle\frac{{\Omega}^x+{\Omega}'^x-i{\Omega}^y{\Omega}'^z+i{\Omega}^z{\Omega}'^y}{1+\hat{\Omega} \cdot \hat{\Omega}'}.\\
    \bra{\hat{\Omega}}{S}^y\ket{\hat{\Omega}'} & = \bra{\hat{\Omega}}\frac{{S}^+ -{S}^-}{2i}\ket{\hat{\Omega}'} = S\bra{\hat{\Omega}}\hat{\Omega}'\rangle\frac{{\Omega}^y+{\Omega}'^y-i{\Omega}^x{\Omega}'^z+i{\Omega}^x{\Omega}'^z}{1+\hat{\Omega} \cdot \hat{\Omega}'}.
\end{align}
Unifying the form of each vector component, we conclude
\begin{equation}
\bra{\hat{\Omega}}\vec{S}\ket{\hat{\Omega}'} = S\bra{\hat{\Omega}}\hat{\Omega}'\rangle\frac{\hat{\Omega}+\hat{\Omega}'-i\hat{\Omega} \times \hat{\Omega}'}{1+\hat{\Omega} \cdot \hat{\Omega}'}.
\end{equation}
and therefore
\begin{equation} 
    \bra{\hat{\bm{\Omega}}}\vec{S}_i\cdot \vec{S}_jU^{\otimes N}\ket{\hat{\bm{\Omega}}} = S^2\vec{\eta}_i(U) \cdot \vec{\eta}_j (U)\bra{\hat{\bm{\Omega}}}U^{\otimes N}\ket{\hat{\bm{\Omega}}},
    \quad \text{where }\vec{\eta}_j(U) = \frac{\hat{\Omega}_j+R(U)\hat{\Omega}_j-i\hat{\Omega}_j \times R(U)\hat{\Omega}_j}{1+\hat{\Omega}_j \cdot R(U)\hat{\Omega}_j}.
\end{equation}
The correlation function then becomes 
\begin{equation}\label{eqn:corr_w_U_int}
\begin{split}
    \mathrm{Tr}\left[\rho_0\vec{S}_i \cdot \vec{S}_j\right] \approx \frac{1}{\mathcal{Z}_0}S^2\int d\hat{\bm{\Omega}}e^{-\beta E[\hat{\bm{\Omega}}]}\int dU \, \vec{\eta}_i(U) \cdot \vec{\eta}_j(U)\bra{\hat{\bm{\Omega}}}U^{\otimes N}\ket{\hat{\bm{\Omega}}}.
\end{split}
\end{equation}

\subsection{Saddle Point Approximation for the Correlation Function}

Below we evaluate the integral over $U$ in the above expression for the correlation function [Eq.~\eqref{eqn:corr_w_U_int}], using the same saddle point approximation method as in Appendix \ref{app:int_over_U}. For simplicity we denote the factor containing the $U$ integral as $K_{ij}(\hat{\bm{\Omega}})$. Substituting Eq.~\eqref{eqn:simplified_integrand} into this integral gives
\begin{align}
    K_{ij}(\hat{\bm{\Omega}}) & \equiv\int dU \vec{\eta}_i(U) \cdot \vec{\eta}_j(U)\bra{\hat{\bm{\Omega}}}U^{\otimes N}\ket{\hat{\bm{\Omega}}}\notag\\
    & = \int d\hat{n}\frac{1}{\pi}\sin^2\frac{\gamma}{2}d\gamma\; \vec{\eta}_i(\hat{\Omega}_i, U) \cdot \vec{\eta}_j(\hat{\Omega}_j, U)\exp\left(2NS \frac{1}{N} \sum_j \log\left(\cos\frac{\gamma}{2}+i\sin\frac{\gamma}{2}\hat{n}\cdot \hat{\Omega}_j\right)\right).
\end{align}
The same approximations used for evaluating $e^{-C(\bm{\hat{\Omega}})}$ can be used here. First, neglecting collinear spin configurations allows us to expand to quadratic order in $\gamma$ around $\gamma = 0$. Second, as before, we can substitute $a \equiv 1-\frac{1}{N}\sum_j (\hat{n}\cdot \hat{\Omega}_j)^2 \approx \frac{2}{3}$ as a constant. These simplifications lead to
\begin{equation}
\begin{split}
    K_{ij}(\hat{\bm{\Omega}}) & \approx \int d\hat{n} \int d\gamma \frac{\gamma^2}{4\pi}\vec{\eta}_i(\hat{\Omega}_i, \gamma = 0) \cdot \vec{\eta}_j(\hat{\Omega}_j, \gamma = 0)\exp\left(2NS \frac{1}{N} \sum_j\left(\frac{1}{8}\left(-1+(\hat{n}\cdot \hat{\Omega}_j)^2\right)\gamma^2+i\frac{\gamma}{2}\hat{n} \cdot \hat{\Omega}_j\right)\right)\\
    & \approx \int d\hat{n} \int d\gamma \frac{\gamma^2}{4\pi}\hat{\Omega}_i \cdot \hat{\Omega}_j \exp\left(2S (-\frac{1}{8} aN \gamma^2+i\frac{\gamma}{2}\hat{n} \cdot \vec{\Omega}_{\text{tot}})\right)
     = \frac{\pi^{-1/2}}{(aSN)^{3/2}}\,\hat{\Omega}_i \cdot \hat{\Omega}_j e^{-\frac{3}{2}\frac{S}{N}\vec{\Omega}_{\text{tot}}^2}
\end{split}
\label{eq:Kij}
\end{equation}

To assess the validity of the expansion in small $\gamma$ to the quadratic order, we consider the leading correction to $K_{ij}$ from expanding the integrand to the next order in $\gamma$:
\begin{equation}\label{eqn:higherW}
\begin{split}
    \delta K_{ij}(\hat{\bm{\Omega}}) & \approx \int d\hat{n} \int d\gamma \frac{-i\gamma^3}{8\pi} (1-\hat{\Omega}_i \cdot \hat{\Omega}_j)(\hat{n} \cdot \hat{\Omega}_i + \hat{n} \cdot \hat{\Omega}_j) \exp\left(2S (-\frac{1}{8}aN \gamma^2+i\frac{\gamma}{2}\hat{n} \cdot \vec{\Omega}_{\text{tot}})\right)\\
\end{split}
\end{equation}
Setting the total $\vec{\Omega}_{\text{tot}}$ as the $\hat{z}$ axis we can write 
\begin{equation} \label{eqn:ndotomega}
\begin{split}
    \hat{n} \cdot \hat{\Omega}_i &= \cos\theta\cos\theta_i + \sin\theta\sin\theta_i\cos(\phi - \phi_i)\\
    & = \cos\theta \hat{\Omega}_i \cdot \vec{\Omega}_{\text{tot}}/|\vec{\Omega}_{\text{tot}}| + \sin\theta\sin\theta_i\cos(\phi- \phi_i)
\end{split}
\end{equation}
The second term in Eq. \eqref{eqn:ndotomega} vanishes upon integration over $\phi$ and the correction becomes
\begin{equation}\label{eqn:corr_2nd_exp}
\begin{split}
    \delta K_{ij}(\hat{\bm{\Omega}}) & = \int d(\cos\theta)\frac{d\phi}{4\pi}\int d\gamma \frac{\gamma^2}{4\pi}\left( -\frac{i\gamma}{2}(1-\hat{\Omega}_i \cdot \hat{\Omega}_j)(\frac{\cos\theta \hat{\Omega}_i \cdot \vec{\Omega}_{\text{tot}}}{|\vec{\Omega}_{\text{tot}}|} + \frac{\cos\theta \hat{\Omega}_j \cdot \vec{\Omega}_{\text{tot}}}{|\vec{\Omega}_{\text{tot}}|})\right)\\
    &\exp\left(2S (-\frac{1}{8}aN \gamma^2+i\frac{\gamma}{2}\cos\theta |\vec{\Omega}_{\text{tot}}|)\right)\\
    & = \int d(\cos\theta) \int d\gamma\frac{-i\gamma^3}{16\pi}\cos\theta \exp\left(2S (-\frac{1}{8}aN \gamma^2+i\frac{\gamma}{2}\cos\theta |\vec{\Omega}_{\text{tot}}|)\right)\left(1-\hat{\Omega}_i \cdot \hat{\Omega}_j\right)\left(\frac{ \hat{\Omega}_i \cdot \vec{\Omega}_{\text{tot}}}{|\vec{\Omega}_{\text{tot}}|} + \frac{\hat{\Omega}_j \cdot \vec{\Omega}_{\text{tot}}}{|\vec{\Omega}_{\text{tot}}|}\right)\\
    & = \frac{\pi^{-1/2}}{a^{5/2}N^{5/2}S^{3/2}}e^{-\frac{3}{2}\frac{S}{N}\vec{\Omega}_{\text{tot}} ^2}(1-\hat{\Omega}_i \cdot \hat{\Omega}_j)(\hat{\Omega}_i + \hat{\Omega}_j )\cdot \vec{\Omega}_{\text{tot}}
\end{split}
\end{equation}
We now estimate how this correction compares to the quadratic order approximation. Denoting by $\braket{\cdots}_0$ the expectation value in the semiclassical theory, and using the approximate singlet constraint $C(\hat{\bm{\Omega}}) \approx \frac{3}{2}\frac{S}{N}\vec{\Omega}_{\text{tot}}^2$, we have
$\langle\vec{\Omega}^2_{\mathrm{tot}}\rangle_0 \sim \frac{N}{S}$. Therefore, $\langle\hat{\Omega}_i\cdot \vec{\Omega}_{\text{tot}}\rangle_0 \sim \langle\vec{\Omega}^2_{\mathrm{tot}}\rangle_0/N\sim 1/S$. So comparing the leading correction to the quadratic order approximation we have $\frac{\delta K_{ij}}{ K_{ij}}\sim\mathcal{O}\left(1/NS\right)$.

Putting it all together, we obtain the correlation function in the form of an average over classical spin configurations:
\begin{equation} \label{corr_approx}
    \mathrm{Tr}\left[\rho_0\vec{S}_i \cdot \vec{S_j}\right]  \approx \frac{1}{\mathcal{Z}}S^2\int d\hat{\bm{\Omega}}\;\hat{\Omega}_i \cdot \hat{\Omega}_j e^{-\frac{3S}{2N} \vec{\Omega}_{\text{tot}}^2} e^{-\beta E[\hat{\bm{\Omega}}]} + \mathcal{O}(S/N).
\end{equation}
This expression is crucial for connecting the subsystem total spin $\vec{\Omega}_A$ in the semiclassical theory to eigenvalues of the subsystem total spin operator.

\section{Gaussian Fluctuation Around Saddle Point}\label{app:Hessian}

The scaling of our lower bound on distillable entanglement and logarithmic negativity in the ordered phase is sensitive to fluctuations about the saddle points. Here we investigate quadratic fluctuations by explicitly diagonalizing the Hessian matrix. This analysis will also allow us to verify that, as expected, the saddle point in Eq.~\eqref{eqn:saddle_ansatz} (and others related by symmetry) is a local minimum of the free energy. 

It is convenient to express fluctuations around the saddle point $\bar{\Omega}(\vec{x})=m(\cos[\vec{K}\cdot \vec{x}],\sin[\vec{K}\cdot \vec{x}],0)$ in terms of a set of position-dependent basis vectors  
\begin{align}
    &\hat{s}_l(\vec{x}) = (\cos[\vec{K}\cdot \vec{x}],\sin[\vec{K}\cdot \vec{x}],0),\notag\\
    &\hat{s}_t(\vec{x}) = (-\sin[\vec{K}\cdot \vec{x}],\cos[\vec{K}\cdot \vec{x}],0),\label{eqn:rotated_basis}\\
    &\hat{s}_z(\vec{x}) = (0,0,1),\notag
\end{align}
such that $\delta\vec{\Omega}(\vec{x}) = \sum_c \omega^c(\vec{x}) \hat{s}_c(\vec{x})$. Writing $\mathcal{S}_0(\vec{\Omega}) = \mathcal{S}_0(\bar{\Omega}) + \delta \mathcal{S}_0(\vec{\Omega})$ we then find that, up to quadratic order in fluctuations,
\begin{align}
    \delta \mathcal{S}^{(2)}_0(\vec{\Omega})= \sum_{\vec{k}}\sum_{c,d=l,t,z} M^{cd} \tilde\omega^c(\vec{k})\tilde{\omega}^d(-\vec{k}) + \ldots,
\end{align}
where $\omega^c(\vec{x})=L^{-d/2}\sum_{\vec{k}}\tilde \omega^c(\vec{k}) e^{-i\vec{k}\cdot \vec{x}}$ and the indices $c$, $d$ label components in the rotated basis defined in Eq.~\eqref{eqn:rotated_basis}. In this basis the Hessian matrix is 

\begin{equation} \label{eqn:Hessian}
\begin{split}
       M  = \begin{pmatrix}
r\vec{k}^2 + 4um^2 & 2ir\vec{k}\cdot \vec{K} & 0\\
-2ir\vec{k}\cdot \vec{K}& r\vec{k}^2 & 0\\
0 & 0 & r\left(\vec{k}^2 - \vec{K}^2\right)
\end{pmatrix} + bS\begin{pmatrix}
\frac{1}{2}\left(\delta(\vec{k}-\vec{K})+\delta(\vec{k}+\vec{K})\right) & \frac{1}{2i}\left(\delta(\vec{k}+\vec{K})-\delta(\vec{k}-\vec{K})\right) & 0\\
\frac{1}{2i} \left(-\delta(\vec{k}+\vec{K})+\delta(\vec{k}-\vec{K})\right) & \frac{1}{2}\left(\delta(\vec{k}-\vec{K})+\delta(\vec{k}+\vec{K})\right) & 0\\
0 & 0 & \delta(\vec{k})
\end{pmatrix}
\end{split}
\end{equation}
The three branches of Hessian eigenvalues are 
\begin{align}
    \lambda_{\pm}(\vec{k}) &= \frac{1}{2}bS \left(\delta(\vec{k} - \vec{K})+\delta(\vec{k} + \vec{K})\right) + r\vec{k}^2 + 2um^2 \notag\\
    & \pm  \frac{1}{2}\sqrt{(bS \delta(\vec{k} - \vec{K})-bS \delta(\vec{k} + \vec{K})+4r\vec{k}\cdot \vec{K})^2+16 u^2 m^4 } 
    \label{eqn:eigval_pm} \\
    \lambda_z(\vec{k}) &= r\left(\vec{k}^2 - \vec{K}^2\right) + bS\delta(\vec{k})\label{eqn:eigval_z}.
\end{align}

To verify that the saddle point $\bar{\Omega}(\vec{x})$ is a stable minimum of the action, note that the $\vec{k}=0$ mode in $\lambda_z$ branch is penalized by $bS$ whenever $S \gg 1/L^2$, and also that $|\vec{K}| \leq |\vec{k}|$ for any nonzero wavevector $\vec{k}$. Therefore $\lambda_z(\vec{k})$ is always non-negative. The $\lambda_{\pm}$ branch is always non-negative provided the mass gap $m \geq |\vec{K}|/\sqrt{u/r} \sim 1/L$, or equivalently $t \leq \sim -1/L^2$, which is the physically relevant regime in the ordered phase.

The Hessian is therefore always positive semi-definite in the physically relevant regime in the ordered phase, so this confirms that the saddle point is a stable local minimum.

Now let us turn to the statistical properties of $\vec{\Omega}_A$. We choose the wave vector $\vec{K}$ to be pointing along the $\hat{x}$ direction. We focus on analyzing the case where $\hat{x}$ is in the plane of the entanglement cut, such that the lower bounds on entanglement measures are sensitive to the fluctuations. We choose the bipartition of the system into $A$ and $B$ to be normal to $\hat{y}$ direction, although the analysis below works for any one of the $d-1$ axis directions orthogonal to $\hat{x}$. In this case $\bar{\Omega}_A = 0$, and by identifying $\vec{\Omega}_A = \sum_{\vec{x} \in A} \delta \vec{\Omega}(\vec{x})$ and summing over the contributions from different fluctuations we find
\begin{equation}\label{eq:mode_sum}
    \langle |\vec{\Omega}_A|^2\rangle = L^d \sum_{n\,\text{odd}}\frac{1}{n^2}\sum_a\langle |\tilde{\omega}^a(n)|^2\rangle,
\end{equation}
where $n$ labels the fluctuation modes $\vec{k} = \frac{2\pi n}{L}\hat{y}$ and the $\sim n^{-2}$ dependence in this expression follows from integrating the half-system spin fluctuations over $A$.

We first analyze spin fluctuations in anisotropic systems in which the minimum nonzero wavevector $\vec{K}$ is unique. We note that with unique $\vec{K}$, the Hessian has exactly three zero modes at $\lambda_-(\vec{k}=0)$ and $\lambda_z(\vec{k}=\pm\vec{K})$, as expected from the fact that this saddle point breaks all three rotational symmetry generators. But as implied in Eqn.~\eqref{eq:mode_sum}, none of these zero modes contributes to $\langle|\vec{\Omega}_A|^2 \rangle$ so we can neglect them. The scaling of each $\langle |\tilde{\omega}^a(n)|^2\rangle$ can be found from the Gaussian Boltzmann weights of these modes [Eqns.\eqref{eqn:eigval_pm}, \eqref{eqn:eigval_z}]: $p[\tilde{\omega}^+ (n)] \sim \exp[-(rL^{-2}n^2 +4um^2)|\tilde{\omega}^+ (n)|^2]$, $p[\tilde{\omega}^- (n)] \sim \exp[-r L^{-2}n^2 |\tilde{\omega}^- (n)|^2]$, and $p[\tilde{\omega}^z (n)] \sim \exp[-r L^{-2}n^2 |\tilde{\omega}^z (n)|^2]$. Substituting the variances into Eq.~\eqref{eq:mode_sum} yields the half-system spin scaling $\langle |\vec{\Omega}_A|^2\rangle \sim r^{-1}L^{d+2}$. Using the same arguments in the main text by matching the moments of $\vec{\Omega}_A^2$ and $s(s+1)$, we get the distribution of $s$ and therefore the scaling of entanglement lower bounds quoted in Eq.~\eqref{eq:twist_para_cut}.

In an isotropic system, however, the minimum nonzero wavevector is not unique. There are $2d$ different $\vec{k}$ such that $|\vec{k}| = |\vec{K}|$. As a consequence, as noted in the main text, a curious feature in the cubic geometry is that $\lambda_z(\vec{k})=0$ for any $\vec{k}\cdot \vec{K}=0$ and $|\vec{k}|=|\vec{K}|$. This gives rise to non-Gaussian fluctuations that seem to contribute to the half-system spin fluctuation, but that do not have a free-energy cost at quadratic order. Such fluctuations are concerning because they have the potential to change the scaling of half-system total spin. 

However, as we now show, expanding beyond quadratic order in fluctuations introduces coupling between these soft fluctuations and the other Hessian eigenvectors. As a consequence, the soft fluctuations acquire a free energy density that is proportional to $L^{-2}$, and do not change the scaling of half-system total spin in situations where the saddle-point configuration has a wavevector $\vec{K}$ in the plane of the bipartition into $A$ and $B$. 

Intuitively, this saddle point configuration should have 3 exact zero modes at $\vec{k} = 0, \pm \vec{K}$, corresponding to the 3 broken generators. Any other zero modes are not protected by symmetry and are purely accidental, so they should be renormalized at higher orders. To see this, we focus on a smooth rotation on $\bar{\Omega}(\vec{x})$ that generates this mode. Consider rotating each $\bar{\Omega}(\vec{x})$ around the axis $\hat{n}(\vec{x}) = (\sin[\vec{K}\cdot \vec{x}],-\cos[\vec{K}\cdot \vec{x}],0)$ for a small angle $\theta(\vec{x})$. The deformed configuration becomes $\vec{\Omega}(\vec{x}) = \cos[\theta(\vec{x})]\bar{\Omega}(\vec{x}) + m\sin[\theta(\vec{x})]\hat{z}$, which generates a $\hat{z}$ component fluctuation. The energy cost of such a fluctuation in the semiclassical theory, at quadratic order in $\theta(\vec{x})$, is 
\begin{equation}
    \delta \mathcal{S}^{(2)}_0 (\theta) \sim \int d^d \vec{x} \, rm^2 \left[(\partial_j \theta)^2-\vec{K}^2 \theta^2\right] +\frac{bS}{L^d}m^2\theta_{\text{tot}}^2= \sum_{\vec{k}}  rm^2 \left(\vec{k}^2-\vec{K}^2 \right)|\tilde{\theta}(\vec{k})|^2 + bSm^2\delta(\vec{k})|\tilde{\theta}(\vec{k})|^2,
\end{equation}
where $\theta(\vec{x}) = L^{-d/2}\sum_{\vec{k}}\tilde{\theta}(\vec{k})e^{-i \vec{k}\cdot \vec{x}}$, confirming that at $\vec{k}^2 = \vec{K}^2$ this is the soft fluctuation that we encountered above. Below we'll denote such a fluctuation, with $|\vec{k}|=|\vec{K}|$ and $\vec{k} \cdot \vec{K}=0$, as $\vec{k}_s$. Now expanding to the quartic order in $\theta(\vec{x})$, we obtain 
\begin{equation}
    \delta \mathcal{S}^{(4)}_0 (\theta) \sim \int d^d \vec{x} \, \frac{1}{3}rm^2\vec{K}^2 \theta^4(\vec{x}) \sim L^{-d-2}\sum_{\delta(\sum\vec{k})}\tilde{\theta}(\vec{k}_1)\tilde{\theta}(\vec{k}_2)\tilde{\theta}(\vec{k}_3)\tilde{\theta}(\vec{k}_4),
\end{equation}
where the $L^{-d-2}$ comes from the $L^{-d/2}$ normalization in the Fourier transform and the $\vec{K}^2 \sim L^{-2}$ factor in the front. Note that the free energy density can depend on $\theta^4(\vec{x})$ (rather than only depending on gradients) because we are considering a rotation around a spatially varying axis. The coupling between different $\vec{k}$ modes then gives an effective mass for the soft fluctuation:
\begin{equation}
    \delta \mathcal{S}^{\text{mass}}_0 (\tilde{\theta}(\vec{k}_s)) \sim L^{-2} \left[L^{-d}\sum_{\text{high }\vec{k}}\langle \tilde{\theta}(\vec{k})\tilde{\theta}(-\vec{k})\rangle\right]|\tilde{\theta}(\vec{k}_s)|^2 \sim L^{-2}|\tilde{\theta}(\vec{k}_s)|^2.
\end{equation}
To the leading order in $L$, this gives the same scaling in half-system spin as quoted in the main text. So we confirmed that such spurious zero mode does not change the statistics of the half-system total spin.

\end{document}